\documentclass[12pt]{article}
\usepackage{amssymb,esvect,amsmath,graphicx,latexsym,amsthm,slashed,eso-pic,hyperref,rotating}

\usepackage{epsfig,amsmath,amssymb,verbatim,mathrsfs,hyperref}
\usepackage{xspace}
\usepackage{xcolor}
\usepackage{slashed}
\usepackage{dcolumn}
\usepackage{multirow}
\graphicspath{{./}{./Figs/}{./figs/}}

\def\beq{\begin{eqnarray}}
\def\eeq{\end{eqnarray}}
\def\bea{\begin{eqnarray}}
\def\eea{\end{eqnarray}}

\def\gev{\, {\rm GeV}}
\def\mev{\, {\rm MeV}}

\newcommand{\gsim}{\lower.7ex\hbox{$\;\stackrel{\textstyle>}{\sim}\;$}}
\newcommand{\lsim}{\lower.7ex\hbox{$\;\stackrel{\textstyle<}{\sim}\;$}}

\newcommand{\nnmb}{\nonumber}
\newcommand{\del}{\partial}

\newcommand{\lrf}[2]{\left(\frac{#1}{#2}\right)}

\newcommand{\kev}{\,\mathrm{keV}}
\newcommand{\ev}{\, {\rm eV}}

\newcommand{\ngam}{\mathcal{N}_{\gamma}}

\newcommand{\eg}{E_{\gamma}}
\newcommand{\ee}{E_{e}}
\newcommand{\nee}{\mathcal{N}_{e}}

\newcommand{\naa}{\mathcal{N}_{a}}
\newcommand{\nbb}{\mathcal{N}_{b}}

\newcommand{\bfgam}{\bar{f}_{\gamma}}
\newcommand{\bfee}{\bar{f}_{e}}
\newcommand{\gamg}{\Gamma_{\gamma}}
\newcommand{\game}{\Gamma_{e}}

\begin{document}

\begin{titlepage}
\noindent
\begin{center}
  \begin{Large}
    \begin{bf}
Limits from BBN on Light Electromagnetic Decays
     \end{bf}
  \end{Large}
\end{center}
\vspace{0.2cm}
\begin{center}
\begin{large}
Lindsay Forestell$^{(a,b)}$, David E. Morrissey$^{(b)}$, and Graham White$^{(b)}$
\end{large}
\vspace{1cm}\\
\begin{it}
(a) Department of Physics and Astronomy, University of British Columbia, Vancouver, BC V6T 1Z1, Canada
\vspace{0.2cm}\\
(b) TRIUMF, 4004 Wesbrook Mall, Vancouver, BC V6T 2A3, Canada
\vspace{0.5cm}\\
email: \emph{\texttt{lmforest@phas.ubc.ca}},
\emph{\texttt{dmorri@triumf.ca}}, 
\emph{\texttt{gwhite@triumf.ca}}
\vspace{0.2cm}
\end{it}
\end{center}
\center{\today}

\begin{abstract}
Injection of electromagnetic energy -- photons, electrons, or positrons --
into the plasma of the early universe can destroy light elements created 
by primordial Big Bang Nucleosynthesis~(BBN).  
The success of BBN at predicting primordial abundances has thus been used 
to impose stringent constraints on decay or annihilation 
processes with primary energies near or above the electroweak scale.  
In this work we investigate the constraints from BBN on electromagnetic
decays that inject lower energies, between 1--100\,MeV.
We compute the electromagnetic cascade from such injections and we
show that it can deviate significantly from the universal spectrum commonly 
used in BBN calculations.  For electron injection below 100\,MeV,
we find that the final state radiation of photons can have a significant
impact on the resulting spectrum relevant for BBN.
We also apply our results on electromagnetic cascades to investigate
the limits from BBN on light electromagnetic decays prior to recombination, 
and we compare them to other bounds on such decays.
\end{abstract}

\end{titlepage}

\setcounter{page}{2}


\section{Introduction\label{sec:intro}}

  Big Bang Nucleosynthesis~(BBN) is one of the most powerful probes
of the very early universe~\cite{Sarkar:1995dd,Iocco:2008va,Jedamzik:2009uy,Pospelov:2010hj}.
Over the course of BBN, free protons and neutrons assemble into a handful 
of light elements~\cite{Schramm:1977ne,Bernstein:1988ad,Walker:1991ap}.  
Assuming a standard $\Lambda$CDM cosmological 
history, the primordial abundances of these elements can be predicted 
using known nuclear reaction rates in terms of a single input parameter, 
the overall baryon density.  
These predictions agree well with observational determinations of 
primordial abundances up to plausible uncertainties in astrophysical 
determinations and nuclear rates~\cite{Cyburt:2015mya}.\footnote{
The extrapolated densities of ${}^6$Li and ${}^7$Li give a particularly
acute puzzle in this regard~\cite{Asplund:2005yt,Sbordone:2010zi,Cyburt:2008kw,Fields:2011zzb}.}

  The success of BBN gives very strong evidence for the $\Lambda$CDM cosmological
model up to radiation temperatures near the MeV 
scale~\cite{Kawasaki:1999na,Kawasaki:2000en,Hannestad:2004px}, 
which extends much earlier than other known tests~\cite{Aghanim:2018eyx}.  
BBN also places stringent constraints on new physics beyond the Standard
Model that injects energy into the cosmological plasma or influences
the expansion rate at early times.  This includes the decays of massive
particles with lifetimes greater than $\tau \simeq 0.1\,\text{s}$~\cite{Ellis:1984er,Juszkiewicz:1985gg,Dimopoulos:1987fz,Reno:1987qw,Dimopoulos:1988ue,Ellis:1990nb,Moroi:1993mb,Kawasaki:1994af,Cyburt:2002uv,Jedamzik:2004er,Kawasaki:2004qu,Jedamzik:2006xz,Kawasaki:2008qe,Kawasaki:2017bqm}, dark matter~(DM) annihilation with an effective cross section near the 
critical value for thermal freeze-out~\cite{Frieman:1989fx,Hisano:2008ti,Hisano:2009rc,Kawasaki:2015yya}, and any new thermalized species with mass below a few MeV~\cite{Cyburt:2004yc,Ho:2012ug,Boehm:2013jpa,Nollett:2013pwa}.

  Limits from BBN on the decays of long-lived massive particles have
been studied in great detail~\cite{Jedamzik:2004er,Kawasaki:2004qu,Jedamzik:2006xz,Kawasaki:2008qe,Kawasaki:2017bqm}.  In the majority of this work,
often motivated by new physics connected to the electroweak hierarchy puzzle
or weakly-interacting massive particle~(WIMP) dark matter, the energy injected 
by the decay has been assumed to be close to or greater than the weak scale.  
Thus, the decay products typically have initial energies that are much
larger than the thresholds for nuclear reactions relevant to BBN which are
typically on the order of several MeV.  Weak-scale decay products typically also 
have both hadronic and electromagnetic components, 
if only through radiative effects.

  Hadronic energy injection can modify the light element abundances at times
as early as $t\sim 0.1\,\text{s}$~\cite{Jedamzik:2004er,Kawasaki:2004qu}.  
Initially, these products scatter with protons and neutrons and alter 
the ratio of these baryons and thus the resulting helium abundance.  
At later times, injected hadrons destroy and modify the abundances 
of helium and other light elements through hadrodissociation.  
Since the initial hadronic energies are usually assumed to be much larger 
than the MeV scale, thresholds for these reactions are easily overcome.

  Electromagnetic energy -- photons, electrons, and positrons -- injected 
into the cosmological plasma does not have a significant effect on the light 
element abundances until much later.  The main effect of electromagnetic
injection on the light elements is photodissociation (unless the amount
of energy deposited is enormous).
However, being much lighter than hadrons, photons and electrons lose their energy
very efficiently by scattering off the highly-abundant photon background.  
The electromagnetic cascade initiated by this scattering is strongly
suppressed for energies above $E_c$, 
given by~\cite{Protheroe:1994dt,Kawasaki:1994sc}
\beq
E_c ~\simeq~ \frac{m_e^2}{22 T} ~\simeq~ (2\,\mev)\lrf{6\,\kev}{T} \ ,
\label{eq:ec}
\eeq
where $T$ is the cosmological photon temperature.
As a result, even for initial energies orders of magnitude above the 
MeV-scale thresholds for photodissociation, the fraction of energy
available for photodissociation is tiny until the background temperature
falls below $T \lesssim 10\,\kev$, corresponding to $t \sim 10^4\,\text{s}$.  

While much of the focus on new sources of energy injection during BBN
has been on decays or annihilations at or above the weak scale, there exist
many well-motivated theories that also predict new sources well below
the weak scale.  Specific examples include dark 
photons~\cite{Fradette:2014sza,Berger:2016vxi}, 
dark Higgs bosons~\cite{Berger:2016vxi,Fradette:2017sdd}, 
dark gluons and glueballs~\cite{Soni:2016gzf,Forestell:2016qhc,Forestell:2017wov}, light or strongly-interacting dark matter~\cite{Henning:2012rm,Hochberg:2018rjs},
and MeV-scale neutrino decays~\cite{Sarkar:1984tt,Ishida:2014wqa}.
As the injection energy falls below the GeV scale, hadronic decay channels
start to become kinematically unavailable and disappear entirely below
the pion threshold.  This leaves electromagnetic and neutrino injection
as the only remaining possibilities.  Even more importantly, it was shown
in Refs.~\cite{Poulin:2015woa,Poulin:2015opa} that the development 
of the electromagnetic cascade at these lower energies can 
differ significantly relative to injection above the weak scale.  
Furthermore, as the injection energy falls below a few tens of MeV, 
photodissociation reactions begin to shut off.

  In this work we expand upon the analysis of 
Refs.~\cite{Poulin:2015woa,Poulin:2015opa} and investigate
the effects of electromagnetic energy injection below $100\,\mev$
on the primordial element abundances created during BBN.  A major focus
of this study is the development of the electromagnetic cascade from
initial photon or electron ($e^+e^-$) injection.  For high energy
injection, the resulting of spectrum of photons is described very well 
by the so-called universal spectrum rescaled by a temperature-
and energy-dependent relaxation rate.  This spectrum is used widely 
in studies of photodissociation effects on BBN, it can be parametrized 
in a simple and convenient way, and has the attractive feature that 
it only depends on the total amount of electromagnetic energy injected.  
However, for lower-energy electromagnetic injection, the universal spectrum 
does not properly describe the resulting electromagnetic cascades.  

The universal spectrum fails for lower-energy injection in two significant ways.
First, the universal spectrum is based on a fast redistribution of the initial 
energy $E_X \gg E_c$ to a spectrum populated at $E \leq E_c$ through Compton 
scattering and photon-photon pair production. 
As shown in Refs.~\cite{Poulin:2015woa,Poulin:2015opa}, this picture does not hold
for initial injection energies $E_X < E_c$, which can easily occur for smaller
$E_X$ and larger decay lifetimes.  And second, as argued 
in Ref.~\cite{Berger:2016vxi} the Compton scattering with background photons 
that dominates electron interactions is qualitatively different at 
high energies compared to low.  At higher energies, 
${s} \sim E\,T \gg m_e^2$, electrons scatter in the Klein-Nishina limit 
and typically lose an order unity fraction of their energy in each 
scattering event.  In contrast, lower energy scattering with 
${s} \sim E\,T \ll m_e^2$ enters the Thomson regime where the fractional 
change in the electron energy per collision is very small and 
the up-scattered photon energy is much less than the initial electron energy.

  To address the breakdown of the universal spectrum for lower-energy
electromagnetic injection, we compute the full electromagnetic cascade
for photon or electron ($e^+e^-$) injection with initial energies
$E_X \in [1,\,100]\,\mev$ following the methods of Ref.~\cite{Kawasaki:1994sc}.
Our work expands upon Refs.~\cite{Poulin:2015woa,Poulin:2015opa} that 
studied photon portion of the cascade for photon injection.  
We compare and contrast our results to the universal spectrum, 
and study their implications for BBN.
In addition to finding important differences from the universal spectrum
at these lower energies, we also demonstrate that final-state radiation~(FSR)
from electron injection can have a very significant impact on the resulting
photon spectrum.  For very low injection energies approaching the MeV scale,
we also study the interplay of the spectrum with the thresholds for the 
most important nuclear photodissociation reactions.

  The outline of this paper is as follows.  After this introduction,
we present our calculation of the electromagnetic cascade
in Sec.~\ref{sec:emc}.  Next, in Sec.~\ref{sec:bbn} we study the 
impact of such electromagnetic injection on the light element abundances.
In Sec.~\ref{sec:other} we contrast the bounds from photodissociation of
light elements with other limits on late electromagnetic injection.
Finally, Sec.~\ref{sec:conc} is reserved for our conclusions.
Some technical details are listed in 
Appendix~\ref{sec:appa} for completeness.

\section{Development of the Electromagnetic Cascade\label{sec:emc}}

  In this section we compute the electromagnetic cascade in the 
early universe following the injection of photons or electrons ($e^+e^-$) 
with initial energy $E_X < 100\,\mev$.  

\subsection{Computing the Electromagnetic Cascade}

  Energetic photons or electrons injected into the cosmological plasma
at temperatures below the MeV scale interact with background photons
and charged particles leading to electromagnetic cascades that produce spectra of 
photons and electrons at lower energies.  Since the development of 
the cascade is much faster than the typical interaction time with
the much more dilute light elements created in BBN, 
these spectra can be used as inputs for the calculation 
of photodissociation effects.

  The most important reactions for the development of the electromagnetic cascade
in the temperature range of interest $T \in [1\,\ev,\,10\,\kev]$ 
are~\cite{Kawasaki:1994sc}:
\begin{itemize}
\item photon photon pair production~(4P): 
$\gamma+\gamma_{BG} \to e^++e^-$
\item photon photon scattering~(PP): 
$\gamma+\gamma_{BG} \to \gamma+\gamma$
\item pair creation on nuclei~(PCN): $\gamma+N_{BG} \to N_{BG}+e^++e^-$
\item Compton scattering~(CS): $\gamma+e^{-}_{BG} \to \gamma+ e^-$
\item inverse Compton~(IC): $e^{\mp}+\gamma_{BG}\to e^{\mp}+\gamma$
\item final state radiation~(FSR): $X \to e^++e^-+\gamma$
\end{itemize}
Of these processes, IC and 4P are typically the fastest provided
there is enough energy for them to occur.

  We define $\mathcal{N}_a = dn_a/dE$ to be the differential number
densities per unit energy of photons ($a=\gamma$) 
and the sum of electrons and positrons ($a = e$).  
The Boltzmann equations for the evolution of these spectra take the form
\beq
\frac{d\naa}{dt}(E) = -\Gamma_a(E)\naa(E) + \mathcal{S}_a(E) \ ,
\eeq
where $\Gamma_a(E)$ is a relaxation rate at energy $E$,
and $\mathcal{S}_a(E)$ describes all sources at this energy.
Since the relaxation rates are typically much faster than the Hubble rate,
the Hubble dilution term has been omitted.  Furthermore, the relaxation
rate is also much smaller than the mean photodissociation rates with light nuclei,
so a further quasistatic approximation can made with 
$d\naa/dt \to 0$~\cite{Kawasaki:1994sc}.
This gives the simple solution
\beq
\naa(E) = \frac{\mathcal{S}_a(E)}{\Gamma_a(E)} \ .
\label{eq:naqs}
\eeq
Note that $\naa(E)$ evolves in time in this approximation
through the time and temperature dependences of the sources and 
relaxation rates.  The source terms are discussed in more detail below
while explicit expressions for the contributions to the relaxation
rates are given in App.~\ref{sec:appa}.

\subsubsection{Monochromatic Photon Injection}
 
For monochromatic photon injection at energy $E_X$ from a decay
with rate per volume $R$, the source terms are
\beq
S_\gamma(E) &=& \xi_\gamma R\;\delta(E-E_X) 
+ \sum_b\!\int_{E}^{E_X}\!\!dE'\,K_{\gamma b}(E,E')\,\nbb(E') \ ,
\label{eq:sgg}
\\
\mathcal{S}_{e}(E) &=& 0 
+ \sum_b\!\int_{E}^{E_X}\!\!dE'\,K_{e b}(E,E')\,\nbb(E') \ ,
\label{eq:seg}
\eeq
where $\xi_{\gamma}$ is the number of photons injected per decay,
and the $K_{ab}(E,E')$ functions describe scattering processes that transfer
energy from species $b$ at energy $E'$ to species $a$ at energy $E \leq E'$.
Explicit expressions for these transfer functions are given 
in App.~\ref{sec:appa}.  Note that in the case of decays of species $X$
with lifetime $\tau_X$, the rate is $R = n_X(t)/\tau_X$.  These
equations can also be applied to annihilation reactions of the form
$X+\bar{X} \to n\,\gamma$ with cross section 
$\langle\sigma v\rangle$ by setting $R = \langle\sigma v\rangle n_Xn_{\bar{X}}$
and $\xi_\gamma = n$.

It is convenient to describe the cascades resulting from the initial
monochromatic (delta function) injection with smooth functions that are
independent of the injection rate.  To this end, we define
\beq
\bfgam(E) &=& \frac{1}{R}\,\ngam(E) 
- \frac{\xi_\gamma}{\Gamma_{\gamma}(E_X)}\,\delta(E-E_X) 
\label{eq:bfgam}\\ 
\bfee(E) &=& \frac{1}{R}\,\nee(E) \ .
\eeq
Using this form in Eq.~\eqref{eq:naqs} with the sources 
of Eqs.~(\ref{eq:sgg},\ref{eq:seg}), we obtain the relations
\beq
\gamg(E)\,\bfgam(E) &=& \xi_\gamma\frac{K_{\gamma\gamma}(E,E_X)}{\gamg(E_X)}
+ \sum_b\!\int_{E}^{E_X}\!\!dE'\,K_{\gamma b}(E,E')\bar{f}_b(E') 
\label{eq:fggeq}\\
\nnmb\\
\game(E)\,\bfee(E) &=&  \xi_\gamma\frac{\,K_{e\gamma}(E,E_X)}{\gamg(E_X)}
+ \sum_b\!\int_{E}^{E_X}\!\!dE'\,K_{e b}(E,E')\bar{f}_b(E')
\label{eq:fegeq}
\eeq
The functions $\bfgam$ and $\bfee$ are expected to be smooth,
and can be used to reconstruct the full spectra $\ngam(E)$
and $\nee(E)$ uniquely for any given injection rate $R$. 

Determining the electromagnetic cascade from monochromatic photon 
injection is therefore equivalent to solving Eqs.~(\ref{eq:fggeq},\ref{eq:fegeq}).
We do so using the iterative method of Ref.~\cite{Kawasaki:1994sc}, 
with an important modification to account for the Thomson limit of IC scattering.
In this method, the spectra $\bar{f}_a(E)$ are determined on a grid
of energy points $E_i$ given by
\beq
E_i = E_0\lrf{E_N}{E_0}^{i/N} \ ,
\eeq
where we use $E_0 = 1\,\mev$, $E_N = E_X$, $i = 0,1,\ldots, N$, and $N\gg 1$.
For the top point $i=N$, Eqs.~(\ref{eq:fggeq},\ref{eq:fegeq}) give
\beq
\bfgam(E_N) = \xi_\gamma K_{\gamma\gamma}(E_N,E_N)/\gamg^2(E_N) \ ,~~~~~~
\bfee(E_N) = \xi_\gamma K_{e\gamma}(E_N,E_N)/\gamg(E_N)\game(E_N) \ .
\eeq
To compute the spectra at lower points, we use the fact that 
the transfer integrals at a given energy $E$ only depend
on the spectra at energies $E' > E$.  Thus, at any step $i$
the integrals in Eqs.~(\ref{eq:fggeq},\ref{eq:fegeq}) can be approximated
numerically (\textit{e.g.} Simpson's rule) using the spectra already 
determined at points $j=i+1,\ldots,N$.  Relative to Ref.~\cite{Kawasaki:1994sc}
we also apply a finer grid to compute the top two energy points.

  This approach to computing the cascades works
well for $y_e = E_eT/m_e^2 \gg 1$, but becomes numerically challenging
for $y_e \lesssim 0.1$.  The problem comes from the contribution
of inverse Compton~(IC) scattering to $K_{ee}$.  As $y_e$ becomes small,
IC scattering enters the Thomson regime in which the cross
section is large but the fractional change in the electron energy
per scattering is much less than unity, and thus the function
$K_{ee}(E,E')$ develops a strong and narrow peak near $E'\simeq E$. 
To handle this we follow Refs.~\cite{Blumenthal:1970gc,Protheroe:1994dt} 
and treat the electron energy loss due to IC in the Thomson limit 
as a continuous process by replacing
\beq
-\Gamma_e(E)\bfee(E) + \int_{E}^{E_N}\!dE'\,K_{ee}(E,E')\bfee(E')
~\to~ \frac{\del}{\del E}\left[\dot{E}\,\bfee(E)\right] \ .
\label{eq:conte}
\eeq
Here, $\dot{E}$ is the rate of energy loss from IC of a single electron
in the photon background, 
given by~\cite{Blumenthal:1970gc}
\beq
\frac{\dot{E}}{E} &=& -\frac{4}{3}\left[\frac{3 \zeta(4)}{\zeta(3)}\right]
\lrf{ET}{m_e^2}\sigma_Tn_{\gamma} \ ,
\eeq
where $\sigma_T = (8\pi/3)\alpha^2/m_e^2$ is the Thomson cross section,
$n_{\gamma} = [2\zeta(3)/\pi^2]T^3$ is the thermal photon density,
and $\zeta(z)$ is the Riemann zeta function.
The approximation of Eq.~\eqref{eq:conte} is valid provided
the fractional energy loss rate $\dot{E}/E$ is much smaller
than the total scattering rate $\sigma_Tn_{\gamma}$,
which coincides with $y_e \ll 0.1$.  In this limit,
the two terms on the left-hand side of Eq.~\eqref{eq:conte} are much
larger than their difference leading to a numerical instability
in the original iterative approach.

When computing the electromagnetic spectra, we use the iterative
method described above with Eqs.~(\ref{eq:fggeq},\ref{eq:fegeq})
until $y_j = E_jT/m_e^2 < 0.05$ is reached.  For lower energy bins
we keep Eq.~\eqref{eq:fggeq} for $\bfgam$ but apply the replacement 
of Eq.~\eqref{eq:conte} for $\bfee$, yielding the solution
\beq
\bfee(E) = \lrf{E_j}{E}^2\bfee(E_j) 
+ \frac{1}{a_TE^2}\,\int_{E}^{E_j}\!dE^{''}\mathcal{S}_e^\prime(E^{''}) \ , 
\eeq
with
\beq
\mathcal{S}_e^\prime(E^{''}) = 
\xi_\gamma\frac{K_{e\gamma}(E^{''}\!,E_N)}{\Gamma_{\gamma}(E_N)}
+ \int_{E^{''}}^{E_X}\!dE'\,K_{e\gamma}(E^{''}\!,E')\,\bfgam(E') \ ,
\eeq
and 
\beq
a_T = \frac{\dot{E}}{E^2} = \frac{4\pi^2}{45}\,\sigma_T\,\frac{T^4}{m_e^2} \ .
\eeq
Again, this can be evaluated iteratively, from high to low.
While we use the specific value $y_e < 0.05$ to match from 
one method to the other, we find nearly identical results from matching 
within the range $y_e \in [0.001,0.1]$.

\subsubsection{Monochromatic Electron Injection}

  Monochromatic injection of electrons (and positrons) at energy $E_X$ 
can be treated nearly identically to monochromatic photon injection,
with the only major change being in modifying the sources to
\beq
S_\gamma(E) &=& S_{\gamma}^{FSR}(E) +
 \sum_b\!\int_{E}^{E_X}\!\!dE'\,K_{\gamma b}(E,E')\,\nbb(E') \ ,
\label{eq:sge}
\\
\mathcal{S}_{e}(E) &=& 
\xi_eR\;\delta(E-E_X) 
+ \sum_b\!\int_{E}^{E_X}\!\!dE'\,K_{e b}(E,E')\,\nbb(E') \ ,
\label{eq:see}
\eeq
where $R$ is the decay (or annihilation) rate per unit volume,
$\xi_e$ is the number of electrons plus positrons injected per decay,
and $S_{\gamma}^{FSR}(E)$ is a contribution to photons from 
final-state radiation to be discussed in more detail below.
For decays of the form $X \to e^++e^-$ we have $R = n_X(t)/\tau_X$
and $\xi_e=2$, while for annihilation $X+\bar{X}\to e^++e^-$ the rate is 
$R = \langle\sigma v\rangle n_Xn_{\bar{X}}$ and $\xi_e=2$.  

  Given these source terms, it natural to define the reduced spectra
$\bar{f}_a(E)$ by
\beq
\bfgam(E) &=& \frac{1}{R}\,\ngam(E) 
\\ 
\bfee(E) &=& \frac{1}{R}\,\nee(E)
- \frac{\xi_e}{\Gamma_{e}(E_X)}\,\delta(E-E_X) 
\eeq
Applying this to Eq.~\eqref{eq:naqs} with the sources 
of Eqs.~(\ref{eq:sge},\ref{eq:see}), we obtain the relations
\beq
\gamg(E)\,\bfgam(E) &=& 
\frac{S_{\gamma}^{FSR}(E)}{R} +
\xi_e\frac{K_{\gamma e}(E,E_X)}{\game(E_X)}
+ \sum_b\!\int_{E}^{E_X}\!\!dE'\,K_{\gamma b}(E,E')\bar{f}_b(E') 
\label{eq:fgeeq}\\
\nnmb\\
\game(E)\,\bfee(E) &=&  \xi_e\frac{\,K_{ee}(E,E_X)}{\game(E_X)}
+ \sum_b\!\int_{E}^{E_X}\!\!dE'\,K_{e b}(E,E')\bar{f}_b(E')
\label{eq:feeeq}
\eeq
These equations can be solved using the same methods as described above
for photon injection, including a matching in the Thomson limit 
using Eq.~\eqref{eq:conte}.

  A new feature that we include for electron injection is a contribution
to the photon spectrum from final-state radiation~(FSR) off the injected
electron; $S_{\gamma}^{FSR}(E)$ in Eq.~\eqref{eq:sge}.  For processes of the
form $X\to e^++e^-$ or $X+\bar{X}\to e^++e^-$ with $X$ uncharged 
and $E_X \gg m_e$, this new source can be approximated 
by~\cite{Birkedal:2005ep,Mardon:2009rc}
\beq
S_{\gamma}^{FSR}(E) \simeq \frac{R}{E_X}\,
\frac{\alpha}{\pi}\,\frac{1+(1-x)^2}{x}\,
\ln\!\left[\frac{4E_X^2(1-x)}{m_e^2}\right]\,
\Theta\!\bigg(1-\frac{m_e^2}{4E_X^2} -x\bigg) \ ,
\eeq
where $x = E/E_X$.
To be fully consistent, a corresponding subtraction should be made from
the electron source.  However, we find that this modifies the spectra by
less than a percent.  In contrast, we show below that the direct contribution
to the photon spectrum from FSR can be the dominant one at
higher energies when $E_XT/m_e^2 \ll 1$, when the initial electrons scatter 
via IC with the photon background mainly in the Thomson regime.

\subsection{Review of the Universal Spectrum}

  Many studies of the effects of electromagnetic energy injection on BBN
approximate the photon spectrum with the so-called universal spectrum.
This is a simple parametrization of the full calculations of the photon spectrum 
in Refs.~\cite{Protheroe:1994dt,Kawasaki:1994sc}.  
It replaces the source terms (direct and cascade) 
in Eq.~\eqref{eq:naqs} with a zeroeth generation spectrum 
$\mathcal{S}_\gamma(E)/R \to p_\gamma(E)$ 
based on the assumption that 4P and IC processes instantaneously reprocess 
the initial injected electromagnetic energy.  

The standard parametrization used for the zeroeth generation spectrum 
is~\cite{Pospelov:2010hj,Cyburt:2002uv,Protheroe:1994dt}
\beq
p_{\gamma}(E_{\gamma}) ~\simeq~ \left\{
\begin{array}{lcl}
0&;&\eg > E_c\\
K_0\lrf{\eg}{E_m}^{-2.0}&;&E_m < \eg < E_c\\
K_0\lrf{\eg}{E_m}^{-1.5}&;&\eg < E_m
\end{array}
\right. \ ,
\eeq
where $E_c \simeq m_e^2/22\,T$ and $E_m \simeq m_e^2/80\,T$ are derived from 
Ref.~\cite{Kawasaki:1994sc}, and $K_0$ is a normalization constant.
For monochromatic injection of $\xi$ photons, electrons, and positrons
each with energy $E_X$, it is fixed by the requirement
\beq
\xi\,E_X = \int_0^{E_X}\!\!\!dE\,E\,p_{\gamma}(E) \ ,
\label{eq:pgnorm}
\eeq
implying $K_0 = \xi E_X/[E_m^2(2+\ln(E_c/E_m)]$ for $E_X > E_c$.  
An important feature of the spectrum is that it is 
proportional to the total injection energy (for either photons or electrons) 
provided $E_X \gg E_c$, up to an overall normalization by the total 
amount of energy injected.  

  Within the universal spectrum approximation, the final spectra are given by
\beq
{f}_{\gamma}(E) = \frac{p_\gamma(E)}{\gamg(E)} 
\ ,~~~~~~~
f_e(E) = 0 \ ,
\eeq
where $f_\gamma(E) = \ngam(E)/R$, and the relaxation rate $\gamg(E)$ accounts 
for the further reprocessing of the spectrum by slower processes 
like Compton scattering, pair creation on nuclei, and photon-photon 
scattering.\footnote{In practice, this $\gamg(E)$ is effectively equal 
to the full relaxation rate that also includes 4P scattering since 
this process is very strongly Boltzmann-suppressed for $E < E_c$.}
These spectra have no residual delta-function parts since the
initial injection is assumed to be fully reprocessed into the zeroeth-order 
spectrum by 4P and IC scatterings.

\subsection{Results for Photon Injection}

\begin{figure}[ttt]
  \begin{center}
    \includegraphics[width = 0.47\textwidth]{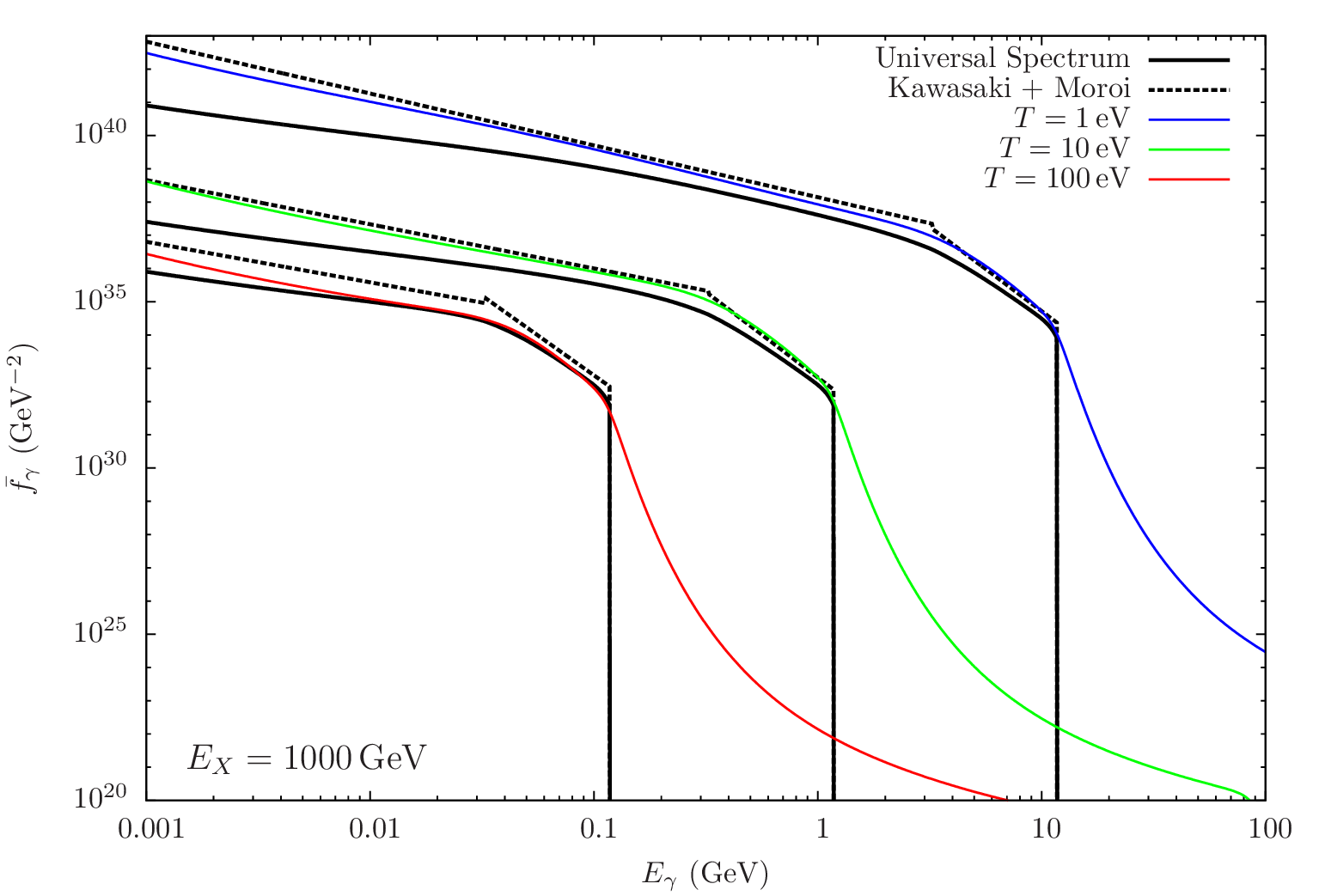}
    \includegraphics[width = 0.47\textwidth]{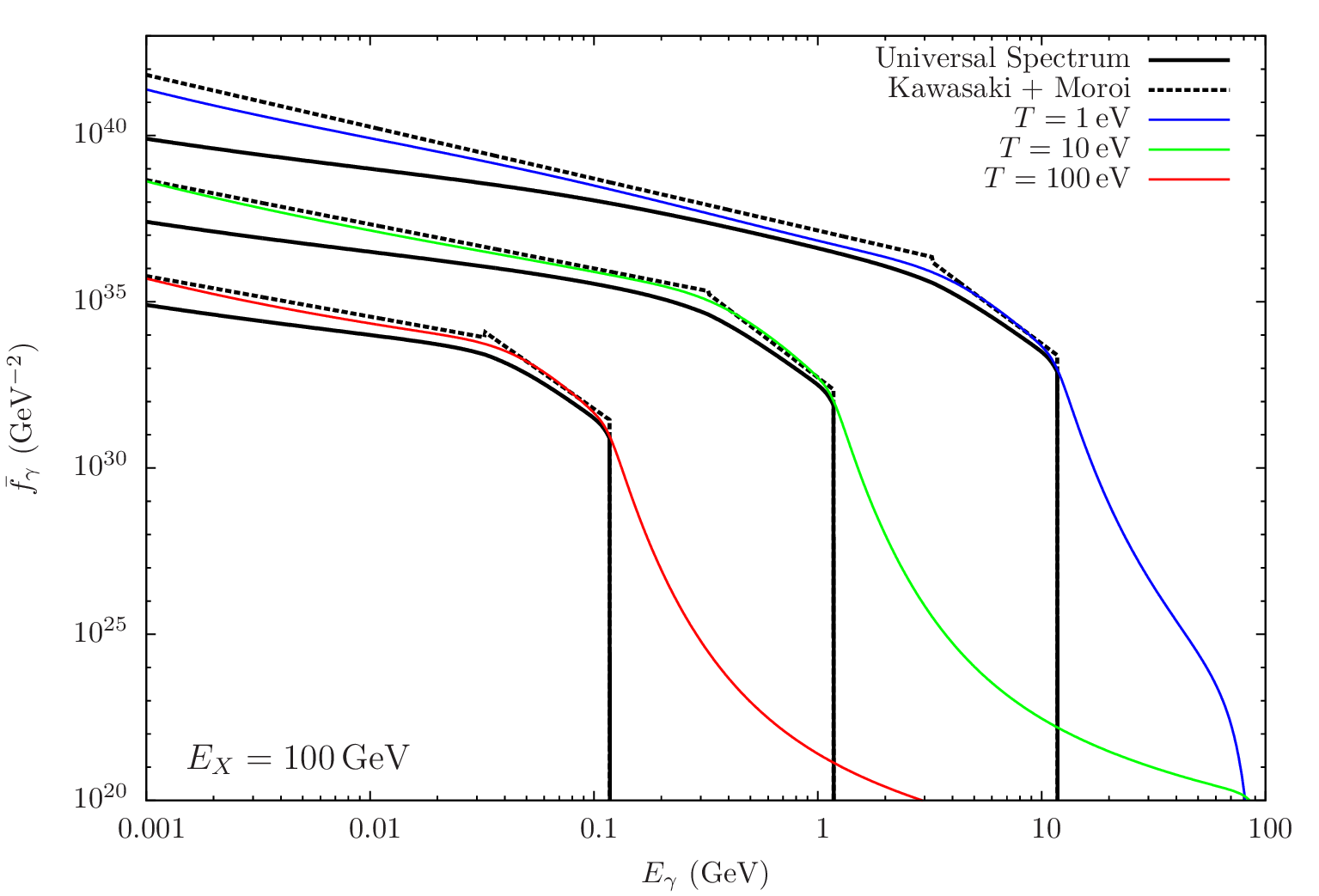}
  \end{center}
\vspace{-0.5cm}
  \caption{Photon spectrum $\bfgam(E)$ for single photon injection 
with energy $E_X=1000\,\gev$~(left) and $100,\gev$~(right), 
for temperatures $T=1,\,10,\,100\,\ev$.  Also shown are the predictions
of the universal spectrum~(solid) and the parametrizations of
Kawasaki and Moroi given in Ref.~\cite{Kawasaki:1994sc}.
}
  \label{fig:ngam23}
\end{figure}

To validate our electromagnetic spectra, we compare our results
to previous calculations and the universal spectrum at high injection
energies.  In Fig.~\ref{fig:ngam23} we show our
photon spectra $\bfgam(E)$ for single photon injection 
with $E_X = 1000\,\gev$~(left) and $100\,\gev$~(right) at temperatures 
$T = 1,\,10,\,100\,\ev$.  
Also shown in the figure are the predictions from the universal spectrum 
and parametrizations of the results of Kawasaki and Moroi listed 
in Ref.~\cite{Kawasaki:1994sc}.  In all cases here, $E_X \gg E_c$
and the universal spectrum is expected to be a good approximation.
Our spectra agree well with the results of Ref.~\cite{Kawasaki:1994sc}
but are somewhat larger than the universal spectrum.  We have also checked
that our spectra scale proportionally to the total energy injected
provided $E_X \gg E_c$.  In all cases shown in the figure, 
the electron spectra are smaller than the photon spectra
by orders of magnitude due to efficient IC scattering.
Also visible is the strong suppression of the photon spectra
for $E > E_c$ where the 4P process is active.

\begin{figure}[ttt]
  \begin{center}
    \includegraphics[width = 0.31\textwidth]{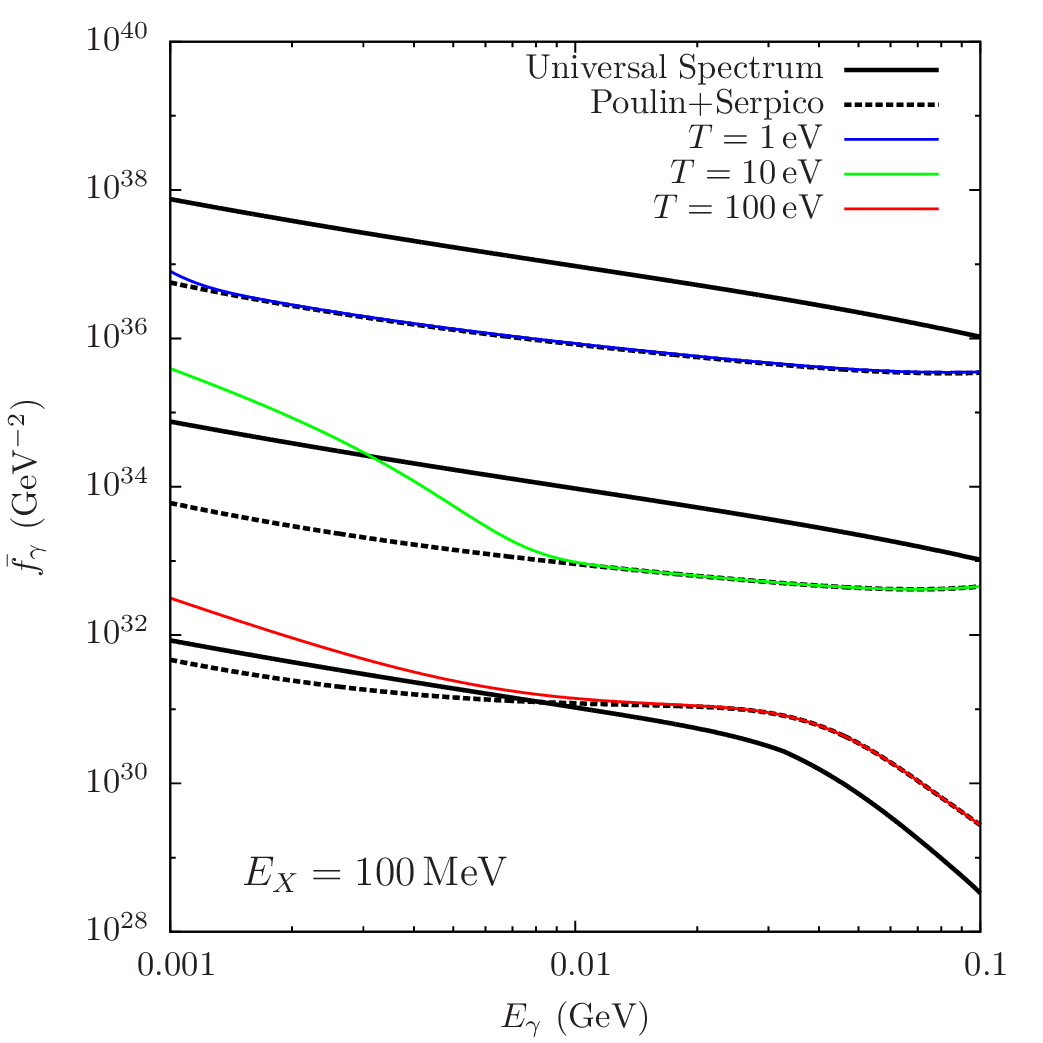}
    \includegraphics[width = 0.31\textwidth]{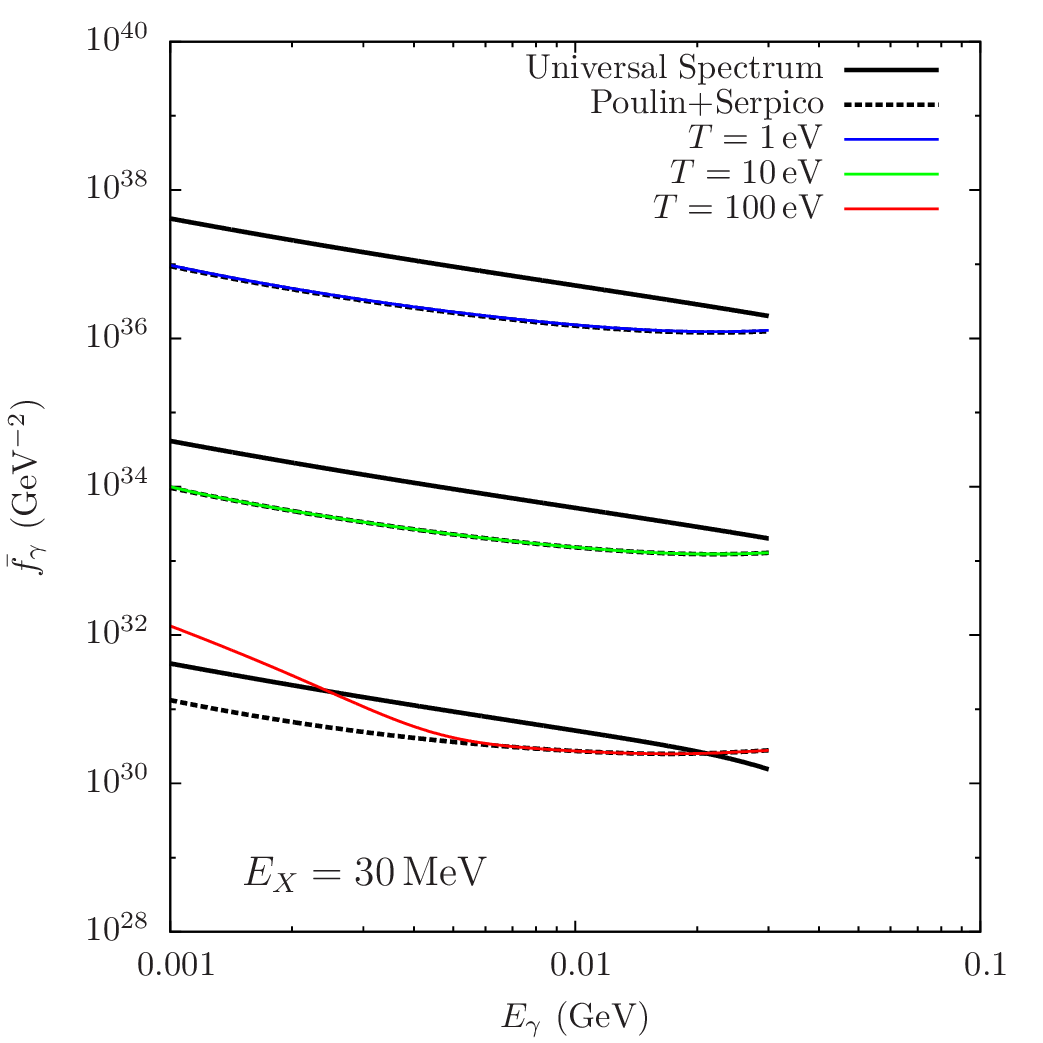}
    \includegraphics[width = 0.31\textwidth]{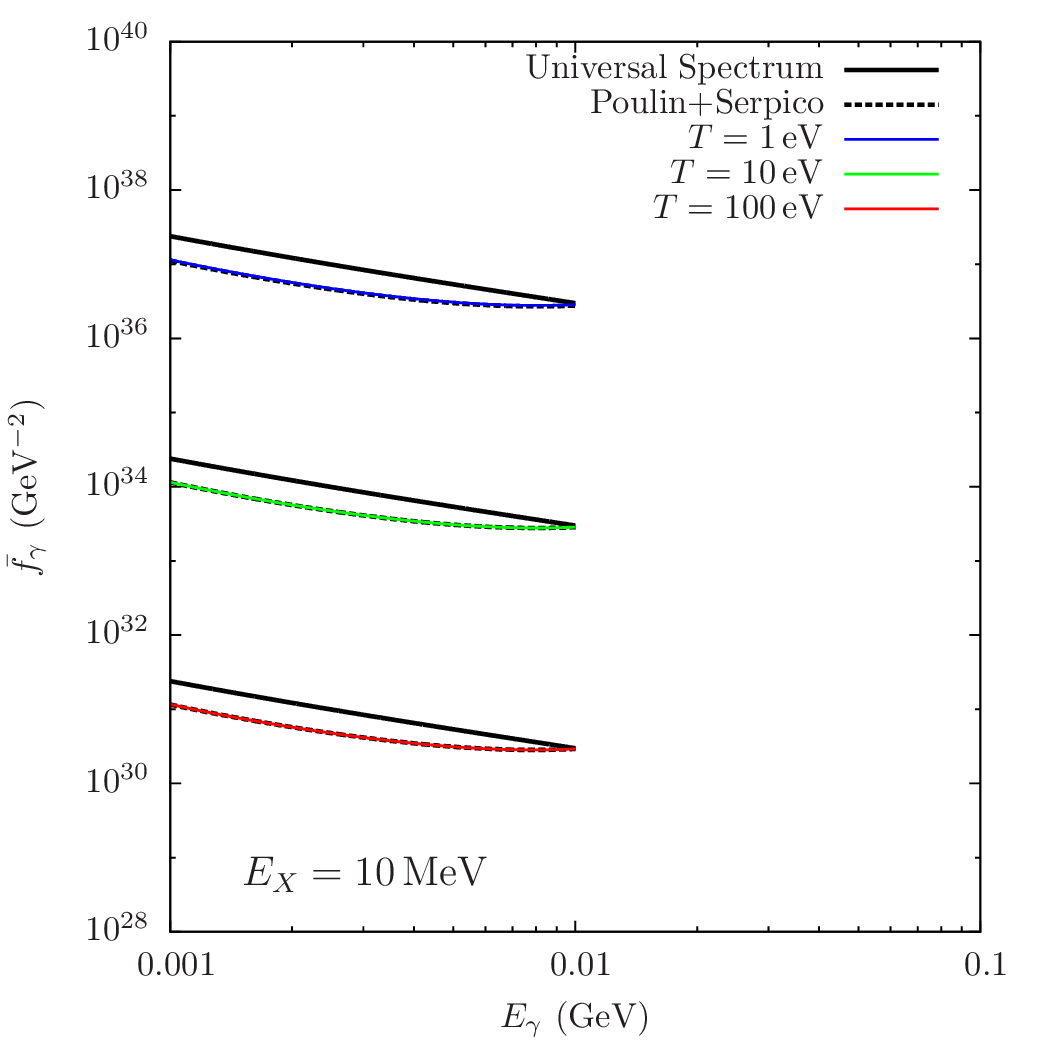}
  \end{center}
\vspace{-0.5cm}
  \caption{Photon spectrum $\bfgam(E)$ for photon injection 
with $E_X=100\,\mev$~(left), $E_X=30\,\mev$~(middle), 
and $E_X=10\,\mev$~(right), with $T=1,\,10,\,100\,\ev$.
Also shown are the predictions of the universal spectrum~(solid) 
and the low-energy prescription of Ref.~\cite{Poulin:2015opa}.
}
  \label{fig:ngam4}
\end{figure}

  In contrast to electromagnetic injection at high energies with $E_X \gg E_c$,
injection at lower energies with $E_X \lesssim E_c$ has received much
less attention.  In Fig.~\ref{fig:ngam4} we show our computed photon spectra 
for single photon injection with $E_X=100\,\mev$~(left), $E_X=30\,\mev$~(middle), 
and $E_X=10\,\mev$~(right) for $T=1,\,10,\,100\,\ev$.
Also shown are the predictions of the universal spectrum (normalized
according to Eq.~\eqref{eq:pgnorm}) and the prescription by Poulin and
Serpico of Ref.~\cite{Poulin:2015opa}.
Since $E_X < E_c$, the assumptions that go into the universal spectrum
are not met and it is not expected to be accurate in this regime,
as first pointed out in Ref.~\cite{Poulin:2015opa}.
Our spectra agree fairly well with the results of Ref.~\cite{Poulin:2015opa},
which only kept the photon part of the spectrum.  Some deviations are seen
at lower energies where photon regeneration by IC becomes significant.
Note as well that the full cascade also contains a moderately damped 
delta-function part that is not shown here (and was explicitly removed
in our definition of $\bfgam$ in Eq.~\eqref{eq:bfgam}).

\subsection{Results for Electron Injection}

\begin{figure}[ttt]
  \begin{center}
    \includegraphics[width = 0.31\textwidth]{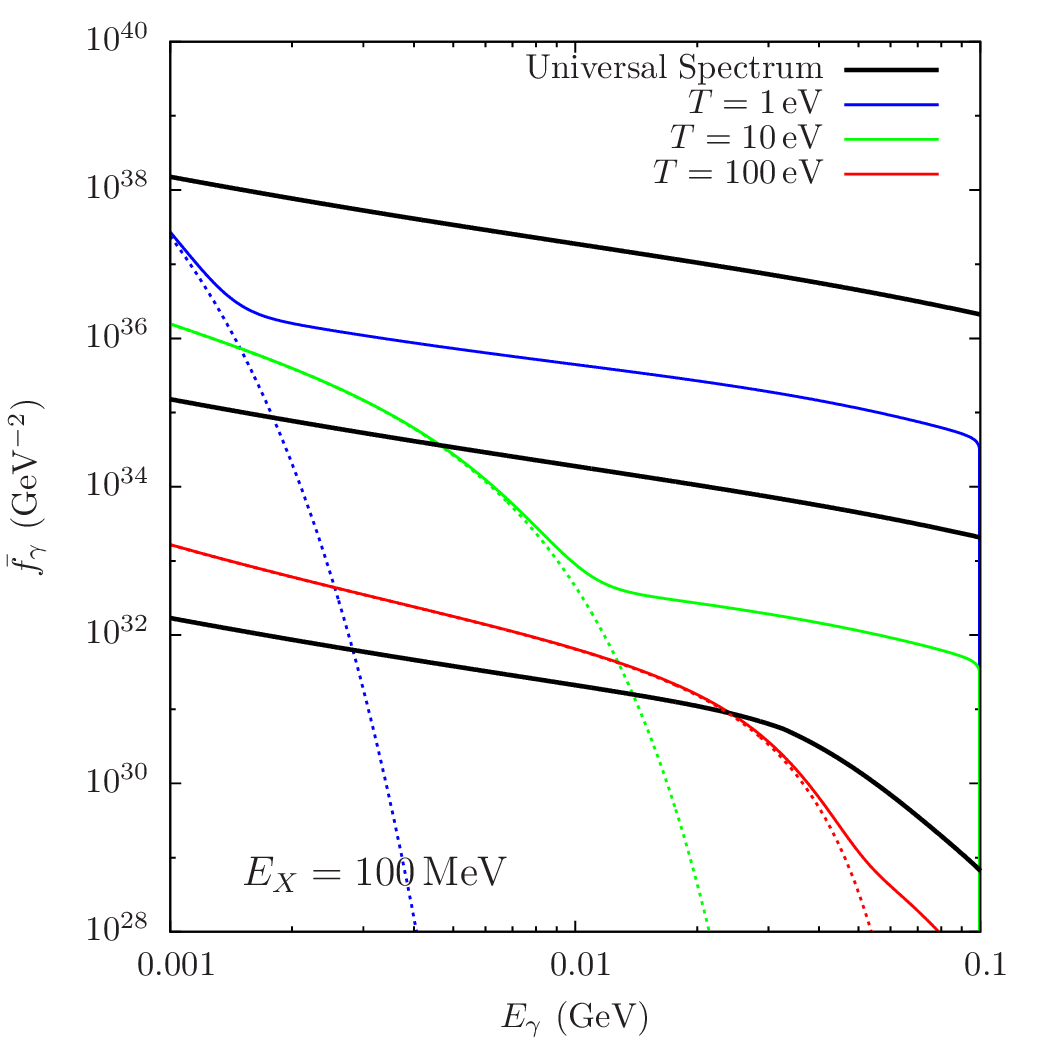}
    \includegraphics[width = 0.31\textwidth]{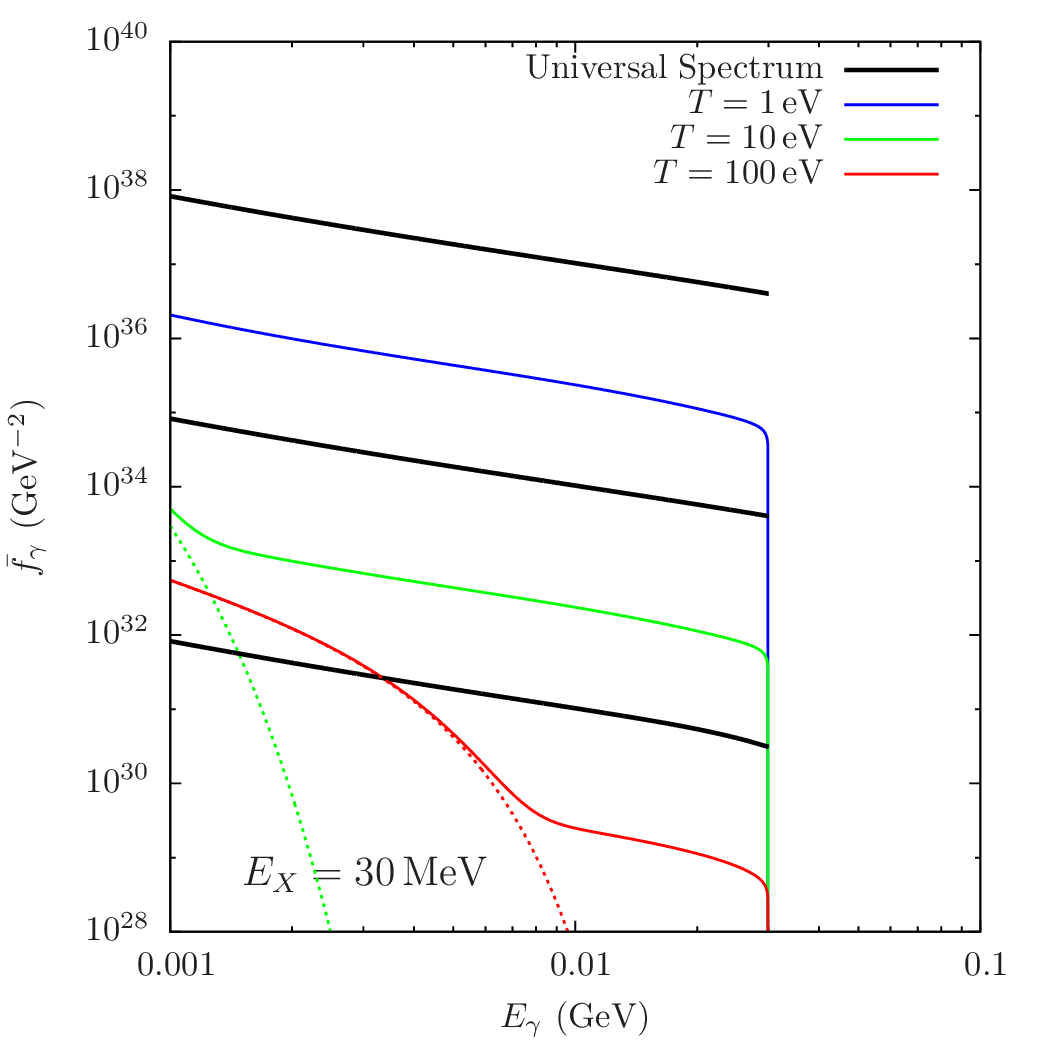}
    \includegraphics[width = 0.31\textwidth]{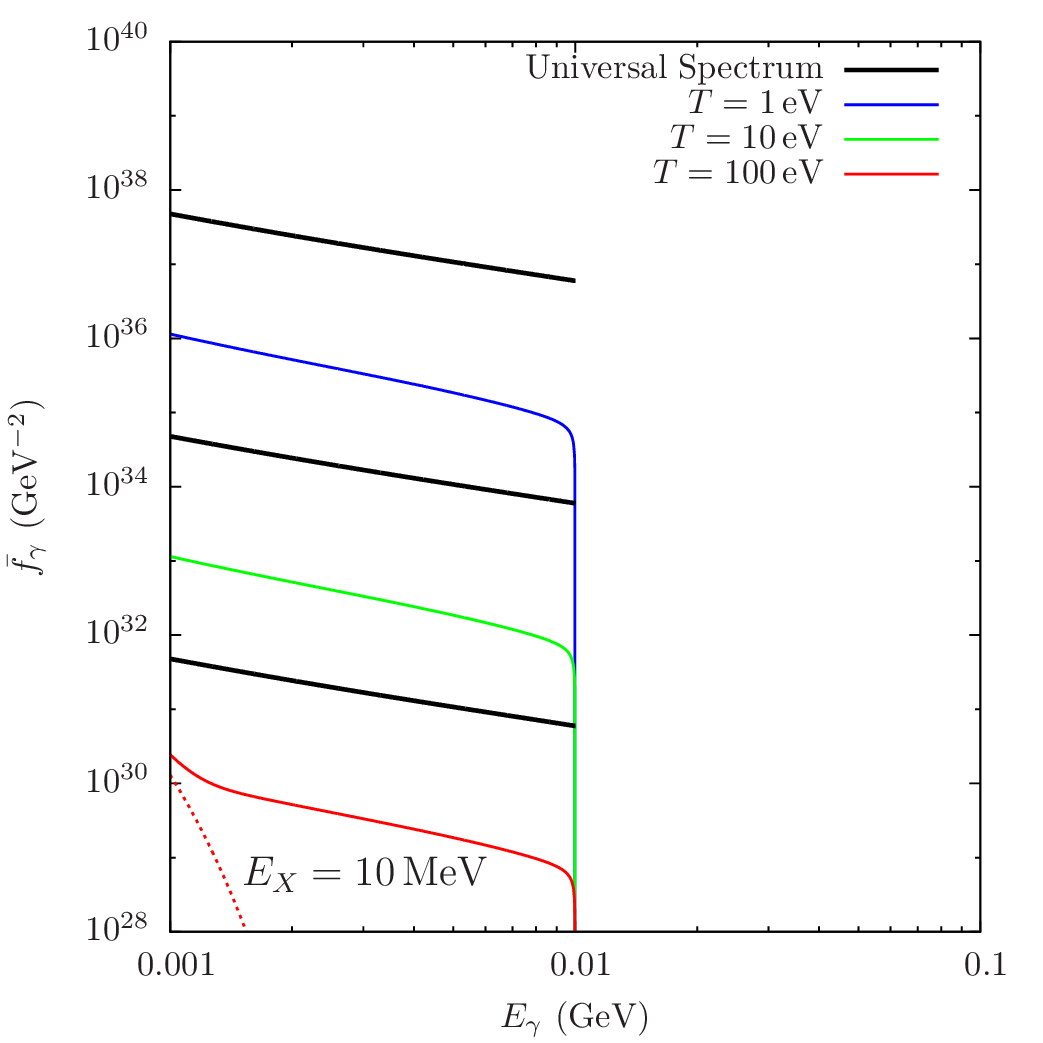}
  \end{center}
\vspace{-0.5cm}
  \caption{Photon spectrum $\bfgam(E)$ for electron plus positron 
($e^+e^-)$ injection with energies $E_X=100\,\mev$~(left), 
$30\,\mev$~(middle), and $10\,\mev$~(right), 
with $T=1,\,10,\,100\,\ev$.  The solid lines show the full spectrum,
while the dashed lines show the result when FSR is not taken into account.
Also shown is the universal spectrum for the same total injected energy.
}
  \label{fig:ngam5}
\end{figure}

  For electron and positron ($e^+e^-$) injection with energies
$E_X \gg E_c$, we find the same photon (and electron) spectra
as from photon injection with an equal total input energy,
and thus our results agree reasonably well with 
Ref.~\cite{Kawasaki:1994sc} and the universal spectrum in this limit.
However, for $E_X \lesssim E_c$ we find very significant variations from
the universal spectrum as well as from pure photon injection.
Photon spectra $\bfgam$ resulting from $e^+e^-$ injection
are shown in Fig.~\ref{fig:ngam5} for input energies
$E_X=100\,\mev$~(left), $E_X=30\,\mev$~(middle), 
and $E_X=10\,\mev$~(right) and temperatures $T=1,\,10,\,100\,\ev$.
The solid lines show the full spectra, while the dashed lines show
the corresponding result when FSR off the initial decay electrons
is not taken into account.  Also shown is the universal spectrum
for the same total energy injection 
(normalized according to Eq.~\eqref{eq:pgnorm}).
Let us also mention that the photon spectra do not have a delta
function component for electron or positron injection.

  The strong suppression of the photon spectrum from electron injection
at lower energies in the absence of FSR was pointed out 
in Ref.~\cite{Berger:2016vxi}.  As argued there, this suppression can
be understood in terms of the behavior of IC scattering at low energy,
which is the main mechanism for electrons to transfer energy to photons
in this context.  For smaller $E_X$ and $T$, the dimensionless combination
$y_e = E_eT/m_e^2 \ll 1$ is small, and IC scattering lies in the 
Thomson regime where each collision only slightly reduces the initial
electron energy.  Correspondingly, the maximal scattered photon energy
$E_{\gamma}'$ in the Thomson limit is $E_{\gamma}^\prime \leq 4(E_e/m_e)E_{\gamma}$,
where $E_{\gamma}$ is the energy of the initial photon.  Since the initial
photon comes from the CMB, $E_\gamma \sim T$ is expected so that  
\beq
E_{\gamma}^\prime &\lesssim& 4(E_e/m_e)T
\label{eq:gcut}\\
&\sim& 15\,\mev\,\lrf{E_e}{100\,\mev}^2\lrf{T}{100\,\ev} \ .\nnmb
\eeq
Higher scattered photon energies are possible, but they come at the 
cost of an exponential Boltzmann suppression.

  In this regime, FSR from the injected electrons and positrons can be
the dominant contribution to the photon spectrum, as illustrated 
in Fig.~\ref{fig:ngam5}.  Relative to the rest of the cascade,
the distribution of photons from FSR is hard, falling off roughly 
as $1/E$ instead of as $1/E^2$.  Despite the suppression of FSR
by $(\alpha/\pi)\times \log$ (with $\log \sim \mathrm{few}$),
it can easily overcome the exponential suppression of IC for 
photon energies above the bound of Eq.~\eqref{eq:gcut}.  
We show below that this has a very important implication for
the effects of lower-energy electron injection on the primordial
light element abundances.  Note, however, that FSR has only a very minor
effect on the spectra for photon injection or when $E_X \gg E_c$.

\section{Effects of Electromagnetic Injection on BBN\label{sec:bbn}}

  Having computed the electromagnetic cascades from lower-energy 
injection, we turn next to investigate the effects of such injection
on the primordial element abundances from BBN.

\subsection{Photodissociation of Light Elements}

\begin{table} \centering \begin{tabular}{lcc}
Process & Threshold (MeV)  & Peak value (mb) \\ \hline
$~\,~\mathrm{D} + \gamma \to p + n$~~\cite{evans2003atomic} & $2.220$ & 2.47  \\ 
${}^3\mathrm{He} +\gamma \to p + p + n$~~\cite{Gorbunov} & 2.486  &1.02\\
${}^3\mathrm{He} + \gamma \to \mathrm{D} + p$~~\cite{Gorbunov} & 5.490 &1.18 \\
$~\,~\mathrm{T} + \gamma \to n + \mathrm{D}$~~\cite{pfiffer:19,Faul:1980zz} & 6.260 &0.818  \\
$~\,~\mathrm{T} + \gamma \to n+n + p$~~\cite{Faul:1980zz} & 8.480 &0.878 \\
${}^4\mathrm{He} +\gamma \to \mathrm{T}+p$~~\cite{Arkatov:1978sq} & 19.81 &1.31  \\
${}^4\mathrm{He} + \gamma \to {}^3\mathrm{He} + n $~~\cite{Irish:1975,Malcom:1973tcd} & 20.58  & 1.28  \\
${}^4\mathrm{He} + \gamma \to \mathrm{D} +\mathrm{D}$~~\cite{Cyburt:2002uv} & 23.85 &0.0051 \\
${}^4\mathrm{He} +\gamma \to n+p+\mathrm{D}$~~\cite{Arkatov:1978sq} &26.07 &0.182
\end{tabular}\label{table:xsections}
\caption{Processes included in our calculation of photodissociation effects from electromagnetic injections, as well as their threshold energies and peak cross sections.}
\end{table}

  Photodissociation of light element begins when the temperature
of the cosmological plasma falls low enough for MeV photons to populate
the electromagnetic cascade.  From Eq.~\eqref{eq:ec},
this does not begin until temperatures fall below about $10\,\kev$
(corresponding to $t\sim 10^4\,\text{s}$).
By this time element creation by BBN has effectively turned off,
and thus we can compute the effects of photodissociation as a
post-processing of the outputs of standard 
BBN~\cite{Jedamzik:2009uy,Cyburt:2002uv}.

  The effects of photodissociation on the light element abundances can
be described by a set of coupled Boltzmann equations of the form
\begin{eqnarray}
\frac{dY_A}{dt} = \sum _i Y_i \int_0^\infty\!d\eg\,  
\ngam(\eg)\,\sigma _{\gamma +i \to A}(\eg) 
- Y_A \sum_f \int_0^{\infty }\!\;d\eg\,\ngam(\eg)\,\sigma_{\gamma + A \to f}(\eg) \ ,
\label{eq:bbnboltz}
\end{eqnarray}
where $\ngam(\eg)$ are the photon spectra calculated above,
$A$ and the sums run over the relevant isotopes, 
and $Y_A$ are number densities normalized to the entropy density,
\beq
Y_A = \frac{n_A}{s} \ .
\eeq
Note that we do not include reactions initiated by electrons because
the electron spectra are always strongly suppressed by IC scattering.

In our analysis we include the nuclear species hydrogen~(H), 
deuterium~($\mathrm{D} = {}^2\mathrm{H}$),
tritium~($\mathrm{T} = {}^3\mathrm{H}$),
helium-3~(${}^3\mathrm{He}$),
and helium~($\mathrm{He} = {}^4\mathrm{He}$).
Heavier species including lithium isotopes could also be included,
but these have much smaller abundances and they would not alter 
the results for the lighter elements we consider.
The nuclear cross sections included in our study are listed 
in Table~\ref{table:xsections}, for which we use the simple parametrizations
of Ref.~\cite{Cyburt:2002uv}.  All these cross sections have the same
general shape as a function of energy, with a sharp rise at the threshold  
up to a peak followed by a smooth fall off.  We list the threshold energies 
and peak values of the cross sections in the table to give an intuitive 
picture of their relevant strengths and ranges of importance.
Of the nine cross sections listed, it is helpful to group them
into processes that destroy helium and create deuterium and helium-3
with thresholds above $20\,\mev$, and processes that destroy the 
lighter isotopes with significantly lower thresholds.

It is straightforward to solve the evolution equations of Eq.~\eqref{eq:bbnboltz}
numerically following the standard convention of converting the dependent variable
from time to redshift.  For standard BBN values of the primordial abundances,
we use the predictions of \texttt{PArthENoPE}~\cite{Pisanti:2007hk,Consiglio:2017pot}:
\beq
Y_p = 0.247 \ ,~~~~~
\frac{n_{\mathrm{D}}}{n_\mathrm{H}} = 2.45\times 10^{-5} \ ,~~~~~
\frac{n_{^3\mathrm{He}}}{n_\mathrm{H}} = 0.998\times 10^{-5} \ .
\eeq
In the analysis to follow, we compare the computed output densities
to the following observed values, quoted with effective $1\sigma$ uncertainties
into which we have combined theoretical and experimental uncertainties 
in quadrature:
\beq
Y_p &=& 0.245\pm 0.004 
\hspace{2.22cm}(\text{Ref.~\cite{Aver:2015iza}})\\
\frac{n_{\mathrm{D}}}{n_\mathrm{H}} &=& (2.53\pm 0.05)\times 10^{-5}
\hspace{1.0cm}(\text{Ref.~\cite{Cooke:2017cwo}}) \\
\frac{n_{^3\mathrm{He}}}{n_\mathrm{H}} &=& (1.0\pm 0.5)\times 10^{-5} 
\hspace{1.4cm}(\text{Ref.~\cite{Geiss:2003ab}})
\ .
\eeq
For the helium mass fraction $Y_p$, the value we use
is consistent with Ref.~\cite{Peimbert:2016bdg} and previous determinations
but significantly lower than the determination of Ref.~\cite{Izotov:2014fga}.
The quoted uncertainty on the ratio $n_{\mathrm{D}}/{n_\mathrm{H}}$ is dominated
by a theory uncertainty on the rate of photon capture on deuterium
from Ref.~\cite{Marcucci:2015yla}.  For $n_{^3\mathrm{He}}/n_\mathrm{H}$,
we use the determination of $(n_{\mathrm{D}}+n_{^3\mathrm{He}})/n_{\mathrm{H}}$
of Ref.~\cite{Geiss:2003ab} together with the value
 of $n_{\mathrm{D}}/{n_\mathrm{H}}$ from Ref.~\cite{Cooke:2017cwo}; 
the resulting upper bound (with uncertainties)
is similar to but slightly stronger than what is used 
in Ref.~\cite{Kawasaki:2017bqm}.  The uncertainties quoted here are
generous, and in the analysis to follow we implement exclusions at
the 2$\sigma$ level.

\subsection{BBN Constraints on Photon Injection}

\begin{figure}[ttt]
    \centering
    \includegraphics[width=0.3\textwidth]{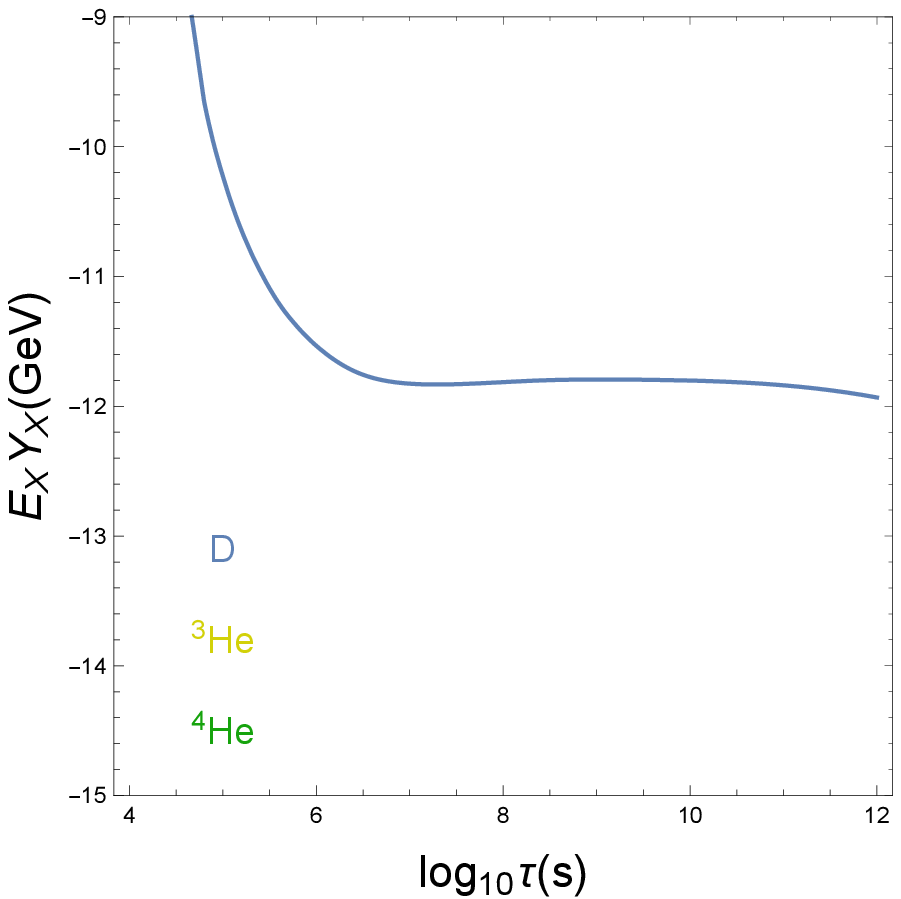} \quad      
    \includegraphics[width=0.3\textwidth]{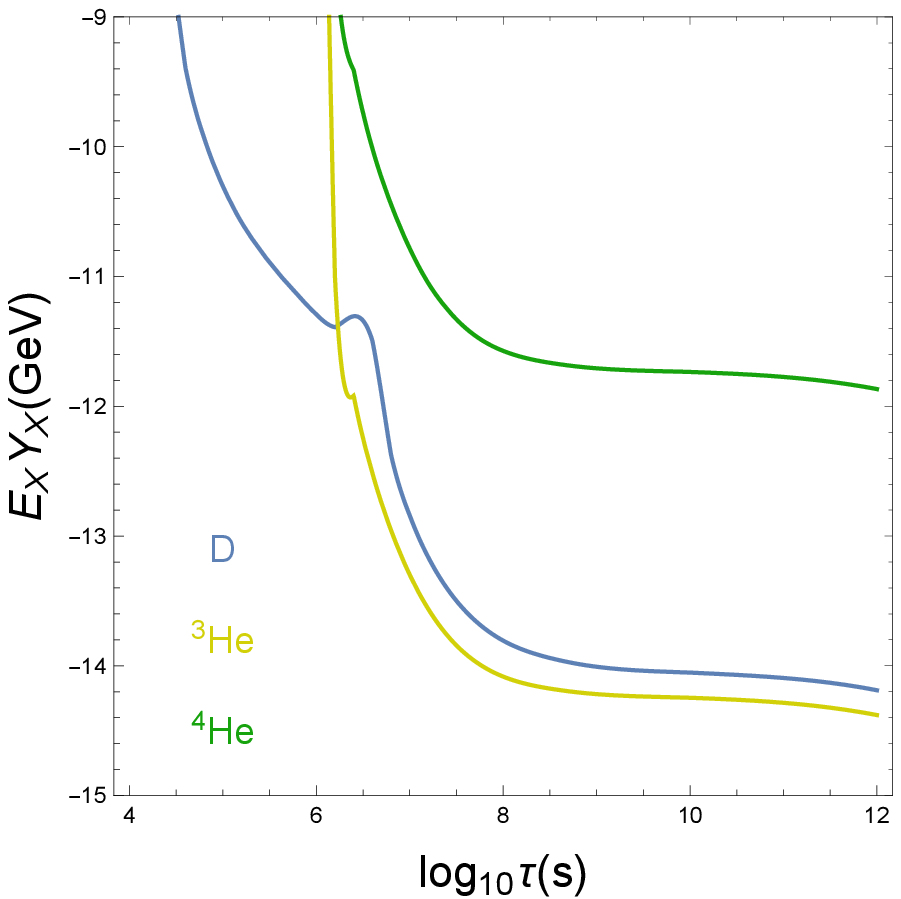} \quad     
    \includegraphics[width=0.3\textwidth]{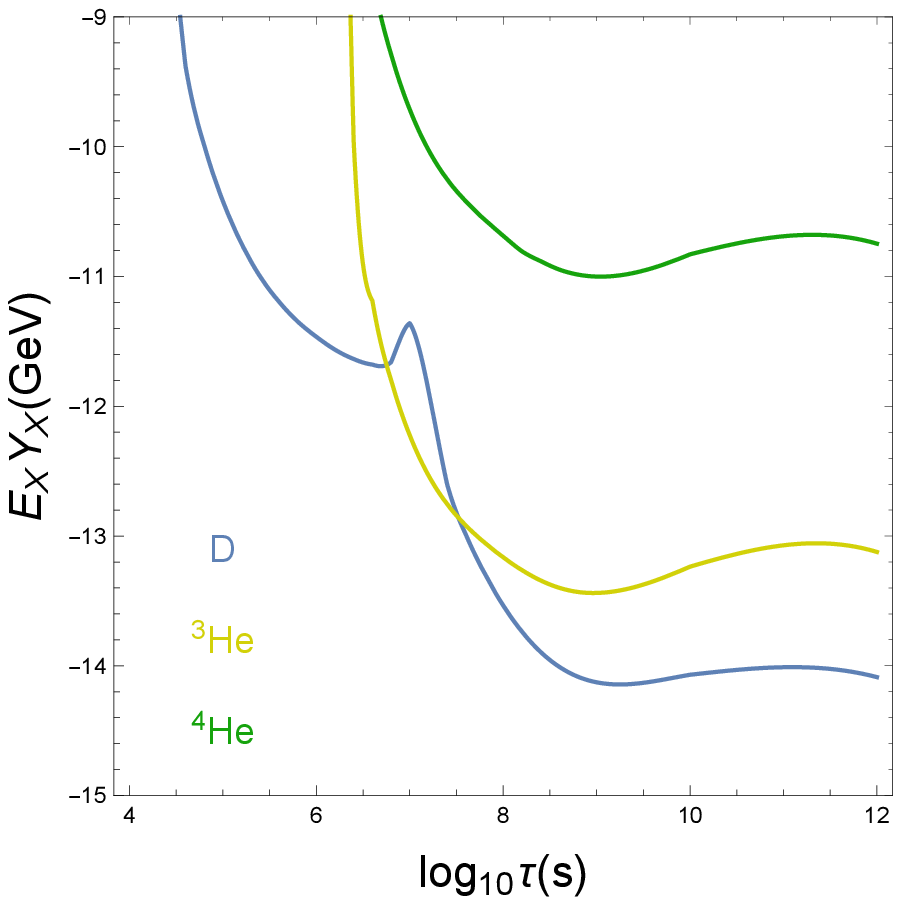}  
    \caption{Limits on $E_X\,Y_X$ from BBN on the monochromatic photon decay 
of species $X$ as a function of the lifetime $\tau_X$ for photon injection
energies $E_X = 10\,\mev$~(left), 30~MeV~(middle), and 100~MeV~(right).
Bounds are given for the effects on the nuclear species $\mathrm{D}$,
${}^3\mathrm{He}$, and ${}^4\mathrm{He}$.}
    \label{fig:ipho}
\end{figure}

  Following the methods described above and the electromagnetic cascades 
computed previously, we derive BBN bounds on monochromatic photon injection 
from late decays with lifetime $\tau_X$ and initial injection energy $E_X$.
In Fig.~\ref{fig:ipho} we show the resulting limits on the combination
$E_X\,Y_X$, where $Y_X$ is the predecay yield of the decaying species
$X$ (assumed to produce one photon per decay) for injection energies
$E_X = 10,\,30,\,100\,\mev$.  The bounds coming from $\mathrm{D}$, 
${}^3\mathrm{He}$, and ${}^4\mathrm{He}$ are shown individually,
and correspond to $2\sigma$ exclusions.  
Early on, when $E_c$ is small, the dominant effect is destruction of 
$\mathrm{D}$ since it has the lowest photodissociation threshold.  
Later on, as $E_c$ increases, it becomes possible to create excess 
$\mathrm{D}$ and ${}^3\mathrm{He}$ through the destruction of ${}^4\mathrm{He}$
provided the injection energy is larger than the ${}^4\mathrm{He}$ threshold
of about $20\,\mev$.  Destruction of $\mathrm{D}$ is the dominant effect
at all times for $E_X$ below the helium threshold, as can be seen in the
leftmost panel of Fig.~\ref{fig:ipho}.

\begin{figure}[ttt]
    \centering 
    \includegraphics[width=0.65\textwidth]{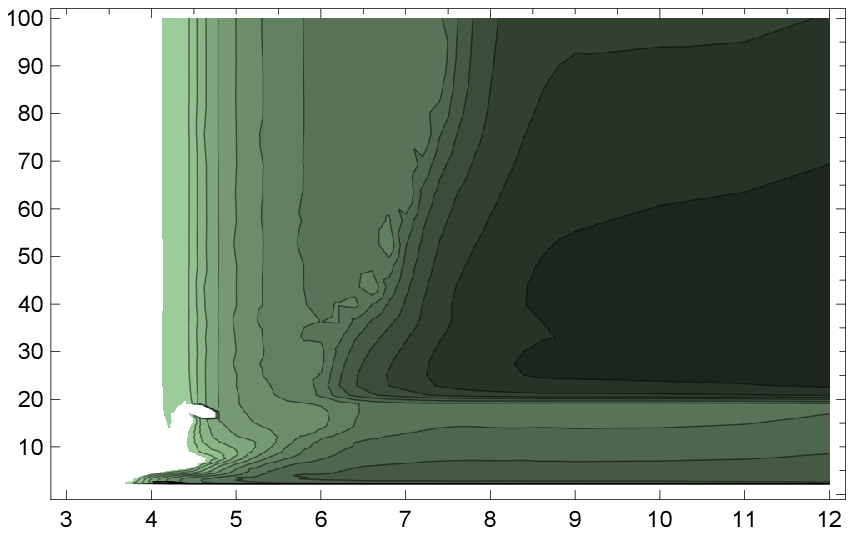} 
\put(-330,100){$\frac{E_X}{\rm MeV}$} 
\put(-150,-20){$\log _{10} \tau (s) $} \quad 
\includegraphics[scale=1.2]{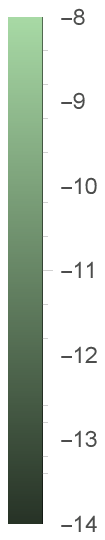} 
\put(-60,-30){$\log _{10} E_X Y_X$~(GeV)}
    \caption{Combined limits on $E_X\,Y_X$ as a function of $\tau_X$ and $E_X$
for the decay of a species $X$ with lifetime $\tau_X$ injecting a single 
photon with energy $E_X$.}
    \label{fig:PHI}
\end{figure}

In Fig.~\ref{fig:PHI} we show maximal values of $E_XY_X$ from monochromatic
photon injection at energy $E_X$ from the decay of species $X$ 
as a function of $\tau_X$ and $E_X$.  The combined exclusion is based 
on the union of $2\sigma$ exclusions of the individual species.
Clear features are visible in this figure at $\tau_X \simeq 10^{6}\,\text{s}$
and $E_X \simeq 20\,\mev$.  These coincide with the structure of the exclusions
shown in Fig.~\ref{fig:ipho}, with both corresponding to where 
the photodissociation of ${}^4\mathrm{He}$ turns off, 
either because $E_c$ or $E_X$ is too small.

\subsection{BBN Constraints on Electron Injection}

\begin{figure}[ttt]
    \centering 
    \includegraphics[width=0.3\textwidth]{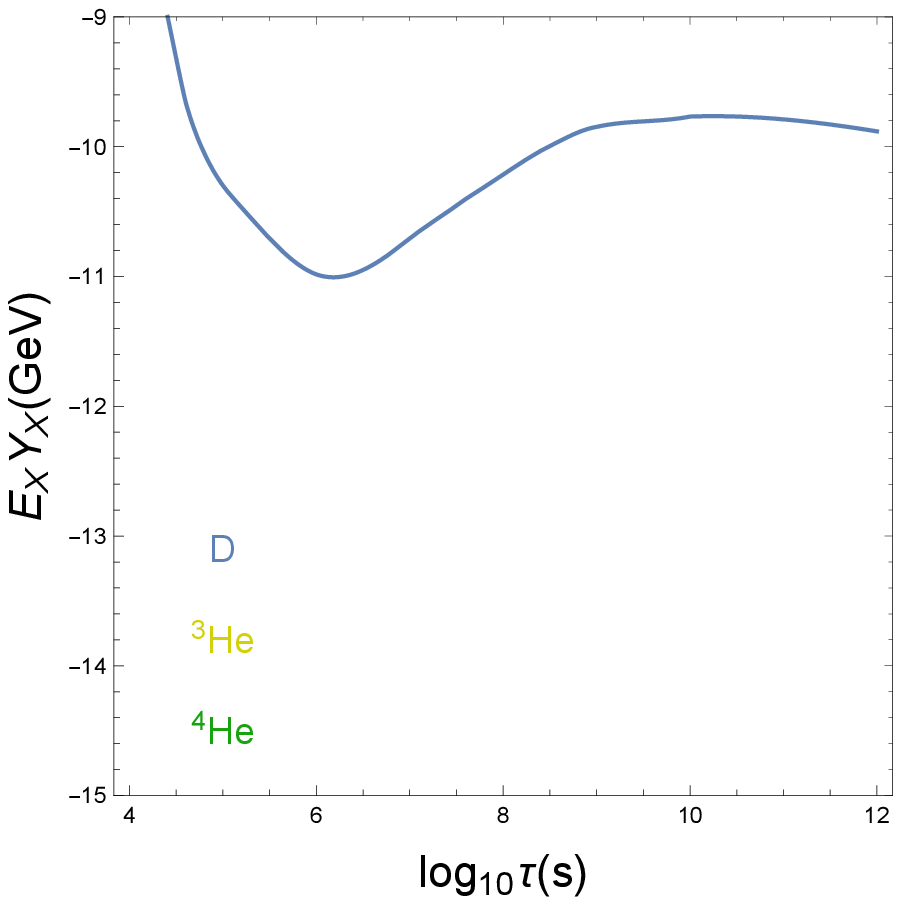} \quad      
    \includegraphics[width=0.3\textwidth]{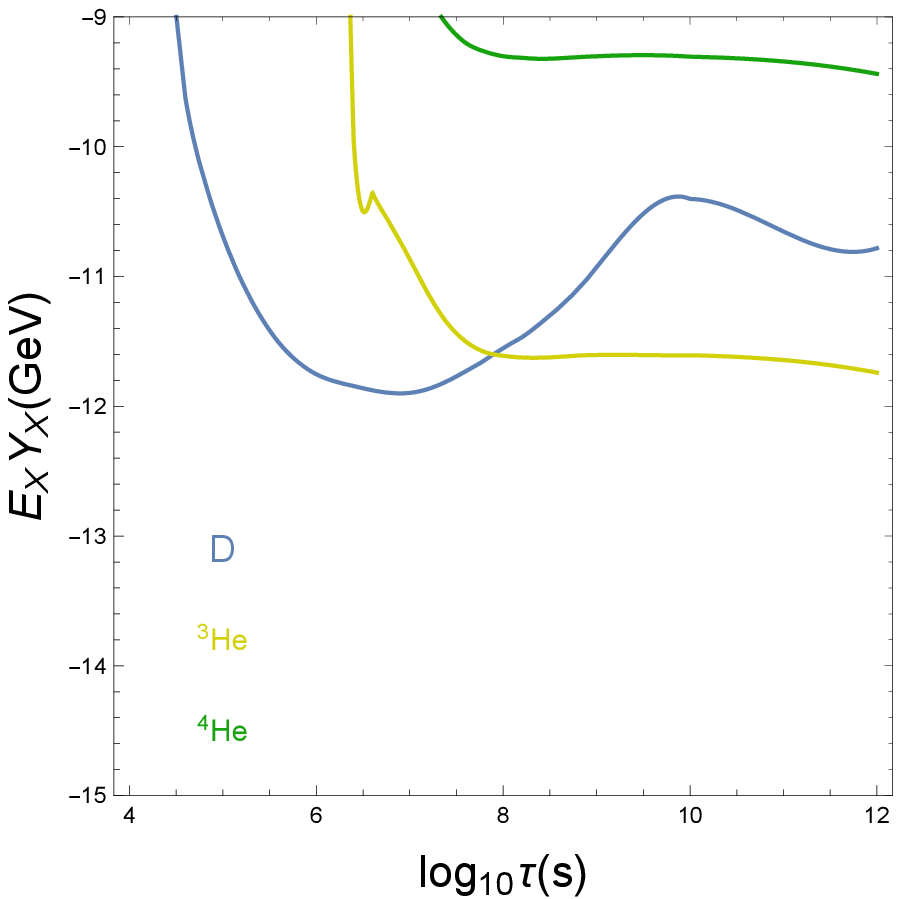} \quad     
    \includegraphics[width=0.3\textwidth]{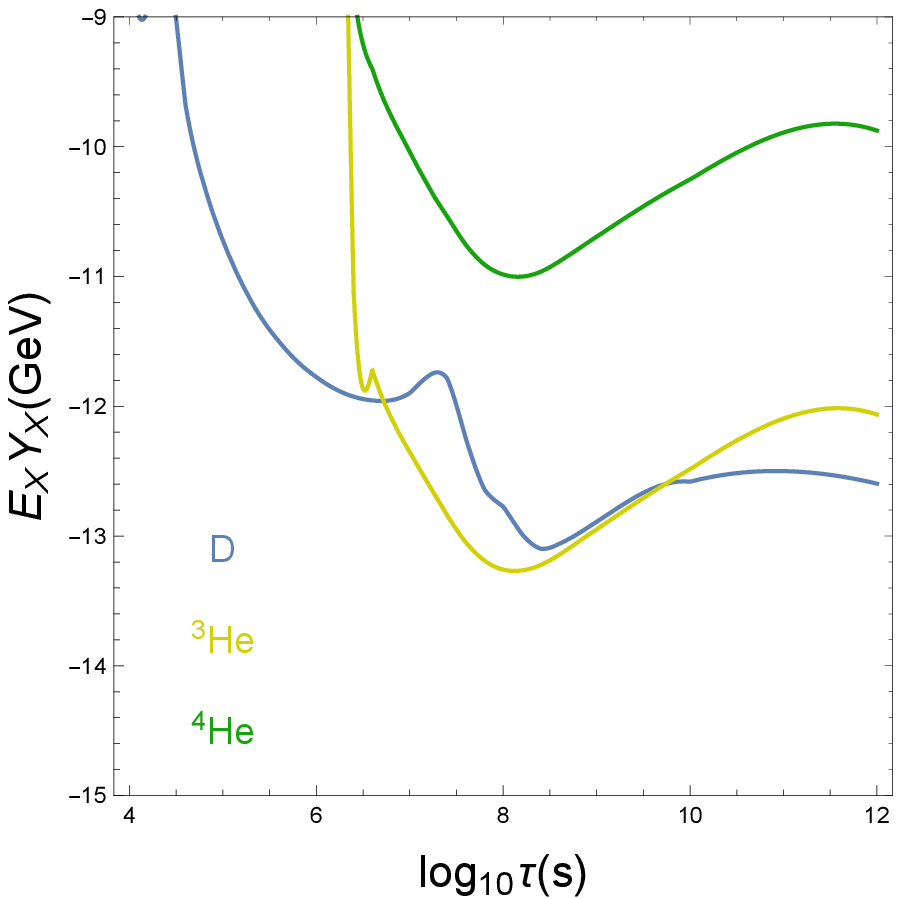}  
    \caption{Limits on $E_X\,Y_X$ from BBN on the monochromatic $e^+e^-$ decay 
of species $X$ as a function of the lifetime $\tau_X$ for individual 
electron injection energies $E_X = 10\,\mev$~(left), 30~MeV~(middle), 
and 100~MeV~(right). Bounds are given for the effects on the nuclear 
species $\mathrm{D}$, ${}^3\mathrm{He}$, and ${}^4\mathrm{He}$,
and contributions to the electromagnetic cascades from FSR are included.}
    \label{fig:ifsr}
\end{figure}

\begin{figure}[ttt]
    \centering
    \includegraphics[width=0.3\textwidth]{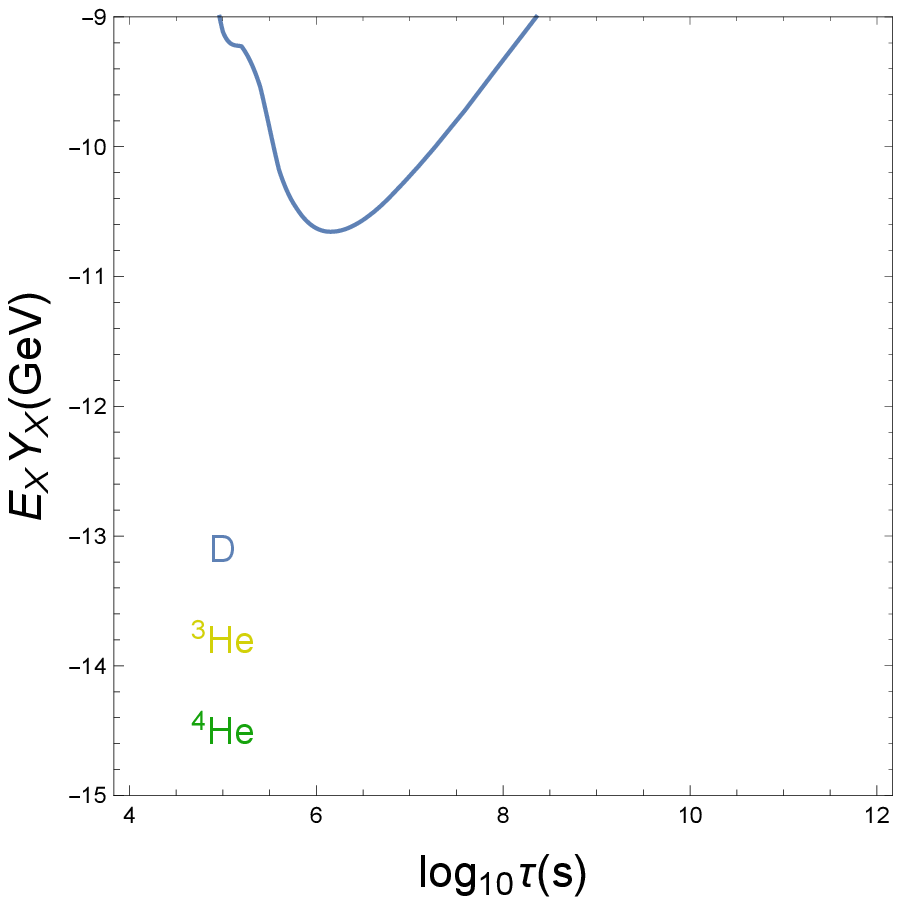} \quad      
    \includegraphics[width=0.3\textwidth]{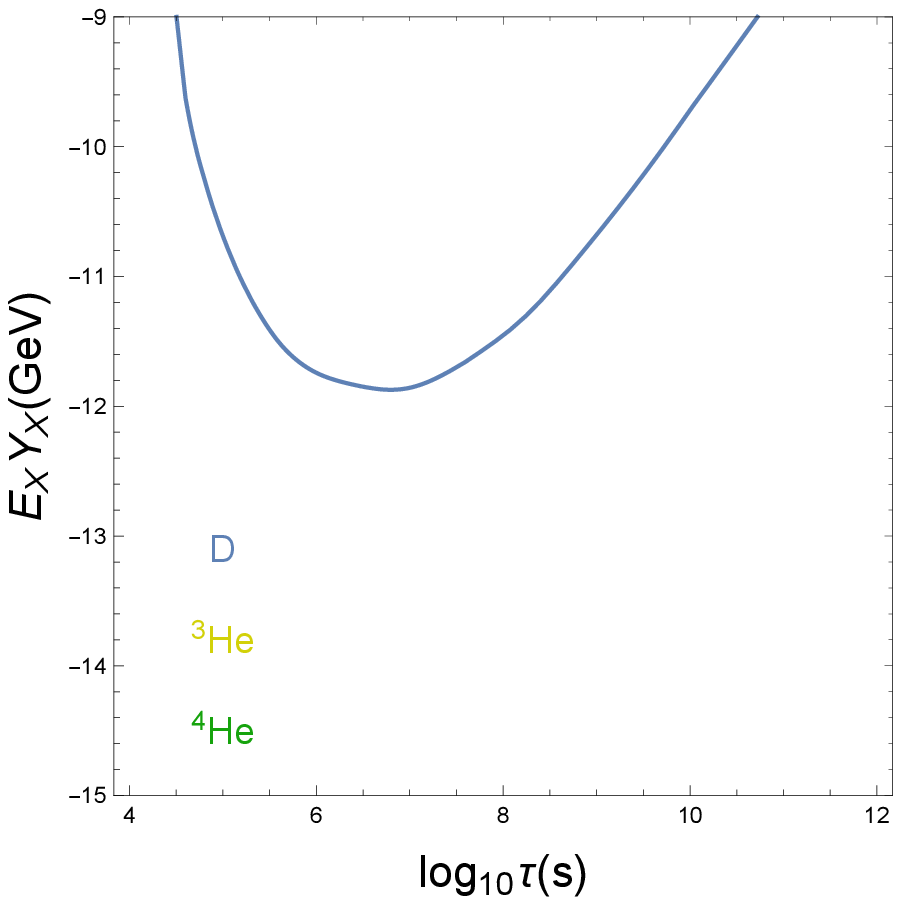} \quad     
    \includegraphics[width=0.3\textwidth]{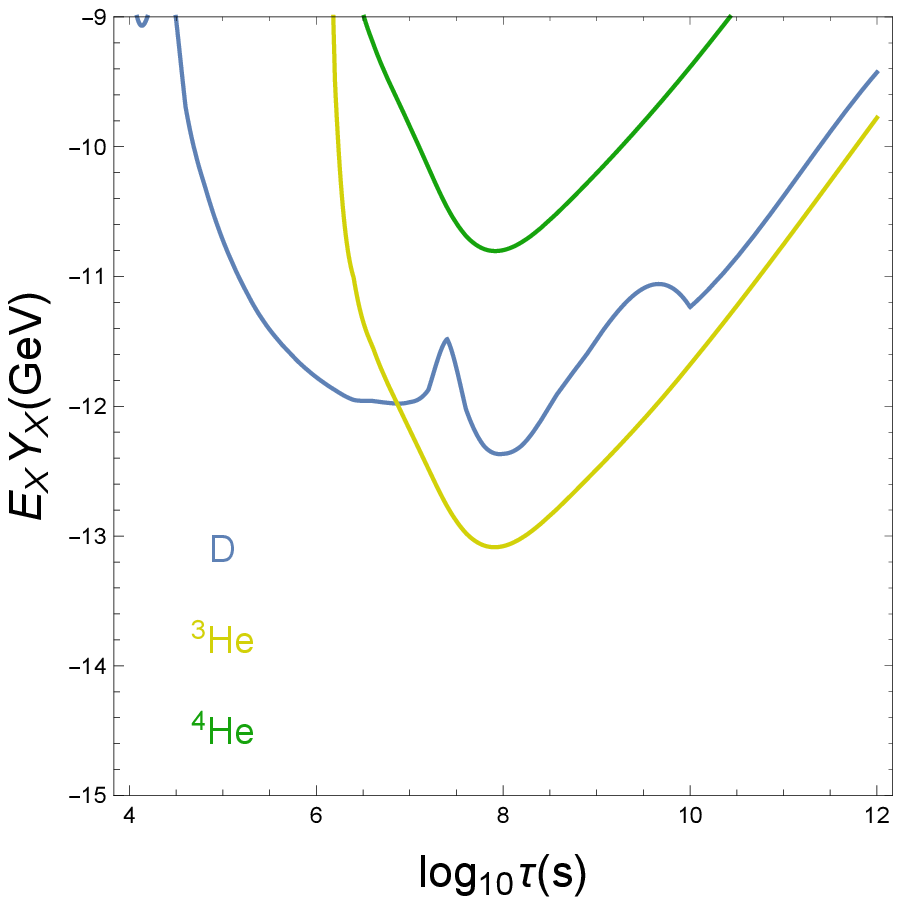}  
    \caption{Same as Fig.~\ref{fig:ifsr} but without FSR effects.}
    \label{fig:ino}
\end{figure}

In Fig.~\ref{fig:ifsr} we show the limits for $e+e^-$ injection
from the decay of a species $X$ with lifetime $\tau_X$
on $E_X\,Y_X$, where $Y_X$ is the predecay yield of the decaying species
$X$ (assumed to produce one $e^+e^-$ pair per decay) for injection energies
for each electron of $E_X = 10,\,30,\,100\,\mev$ (from left to right).  
The bounds coming from $\mathrm{D}$, ${}^3\mathrm{He}$, and ${}^4\mathrm{He}$ 
are shown individually, and correspond to $2\sigma$ exclusions.  
The electromagnetic spectra used in this calculation include FSR from 
the injected $e^+e^-$ pair.  The resulting bounds are somewhat weaker 
than for photon injection and follow a similar pattern, 
and remain quite strong even down to $E_X = 10\,\mev$.
For comparison, we show the corresponding results when FSR effects
are not included in Fig.~\ref{fig:ino}.  
As expected, the exclusions are significantly weaker,
particularly for larger $\tau_X$ and lower $E_X$ 
where the relevant IC scattering is deep in the Thomson regime.  

\begin{figure}[ttt]
    \centering 
    \includegraphics[width=0.65\textwidth]{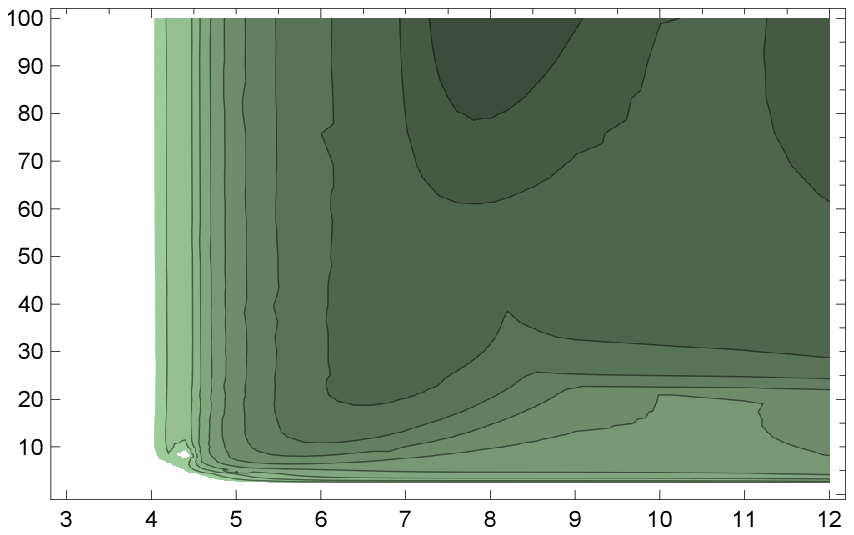} 
\put(-330,100){$\frac{E_X}{\rm MeV}$} 
\put(-150,-20){$\log _{10} \tau (s) $} \quad 
\includegraphics[scale=1.2]{legendall.eps} 
\put(-60,-30){$\log _{10} E_X Y_X$~(GeV)}
    \caption{Combined limits on $E_X\,Y_X$ as a function of $\tau_X$ and $E_X$
for the decay of a species $X$ with lifetime $\tau_X$ injecting an 
electron-positron pair each with energy $E_X$, with FSR effects included.}
    \label{fig:fsr}
\end{figure}

In Fig.~\ref{fig:fsr} we show maximal values of $E_XY_X$ from monochromatic
$e^+e^-$ injection at energy $E_X$ from the decay of species $X$ 
as a function of $\tau_X$ and $E_X$, with FSR effects included in 
the electromagnetic cascade.  The combined exclusion is based 
on the union of $2\sigma$ exclusions of the individual species.
Again, the exclusions become weaker for $\tau_X \lesssim 10^{6}\,\text{s}$
or $E_X \lesssim 20\,\mev$ where the photodissociation of ${}^4\mathrm{He}$ 
turns off.  The bounds on $e^+e^-$ injection are also typically weaker
than for photon injection, but not drastically so when FSR is taken into account.

\section{Other Constraints on Electromagnetic Decays\label{sec:other}}

  In addition to modifying the primordial light element abundances,
energy injection in the early universe can produce other deviations
from the standard cosmology.  Electromagnetic decays near or after recombination 
at $t_{rec} \simeq 1.2\times 10^{13}\,\text{s}$ can modify the the temperature 
and polarization power spectra of the CMB~\cite{Adams:1998nr,Chen:2003gz,Padmanabhan:2005es,Zhang:2007zzh}.  
Since current CMB observations are found to constrain such decays much 
more strongly than BBN~\cite{Slatyer:2009yq,Cline:2013fm}, 
we focus here on decays prior to recombination.
The best limits in this case, aside from BBN, come from entropy injection 
and modifications to the CMB frequency spectrum.
In this section we estimate these other limits on late energy injection
and compare them to our results for BBN.

  Entropy injection after the start of BBN leads to a lower
measured baryon density today relative to the value deduced from BBN.
This was studied in Ref.~\cite{Feng:2003uy} with the result
\beq
\frac{\Delta s}{s} ~\simeq~ 7.8\times 10^{-5}
\lrf{\Delta E\,Y_X}{10^{-10}\,\gev}
\bigg(\frac{\tau}{10^6\,\text{s}}\bigg)^{1/2} \ ,
\eeq
where $\Delta E$ is the total electromagnetic energy injected
per decay and $\tau_X \gtrsim 1\,\text{s}$.  A related constraint
can be derived for variations in the effective number of neutrinos $N_{eff}$
from photon heating after neutrino decoupling~\cite{Boehm:2013jpa,Nollett:2013pwa,Ishida:2014wqa,Kaplinghat:2000jj}.

    Late decays releasing electromagnetic energy can also distort the
frequency spectrum of the CMB~\cite{Hu:1992dc,Hu:1993gc}, 
which is observed to be a nearly-perfect blackbody~\cite{Fixsen:1996nj}.
The effect depends on the decay time $\tau_X$ relative
to the times $\tau_{dC} \simeq 6.1\times 10^{6}\,\text{s}$ 
when double-Compton scattering freezes out 
and $\tau_C \simeq 8.8\times 10^{9}\,\text{s}$ 
when Compton scattering turns off~\cite{Hu:1992dc,Hu:1993gc}.
%
%
Decays with $\tau_{dC} < \tau_X < \tau_C$ yield products that thermalize
through Compton scattering and generate an effective photon chemical potential 
$\mu$ given by~\cite{Hu:1992dc,Hu:1993gc,Chluba:2011hw}
\beq
\mu ~\simeq~ 5.6\times 10^{-4}
\lrf{\Delta E\,Y_{X}}{10^{-10}\,\gev}
\bigg(\frac{\tau}{10^6\,\text{s}}\bigg)^{1/2}
e^{-(\tau_{dC}/\tau)^{5/4}} \ .
\eeq
For $\tau_X > \tau_C$, electromagnetic injection produces a distortion 
that can be described by the Compton parameter 
$y = \Delta\rho_{\gamma}/4\rho_{\gamma}$,
with the approximate result~\cite{Hu:1992dc,Hu:1993gc,Chluba:2011hw}
\beq
y ~\simeq~ 5.7\times 10^{-5}
\lrf{\Delta E\,Y_{X}}{10^{-10}\,\gev} 
\bigg(\frac{\tau}{10^6\,\text{s}}\bigg)^{1/2}\!\mathcal{C}(\tau) \ ,
\eeq
where $\mathcal{C}(\tau) = 1$ for $\tau < t_{eq}$ 
and $\mathcal{C}(\tau) \simeq (\tau/t_{eq})^{1/6}$ for $\tau > t_{eq}$.
The current limits on $\mu$ and $y$ are~\cite{Fixsen:1996nj}
\beq
\mu < 9\times 10^{-5},~~~~~|y| < 1.5\times 10^{-5} \ ,
\eeq
while the proposed PIXIE satellite is to have sensitivity 
to constrain~\cite{Kogut:2011xw}
\beq
\mu < 1\times 10^{-8},~~~~~|y| < 2\times 10^{-9} \ .
\eeq

\begin{figure}[ttt]
  \begin{center}
    \includegraphics[width = 0.45\textwidth]{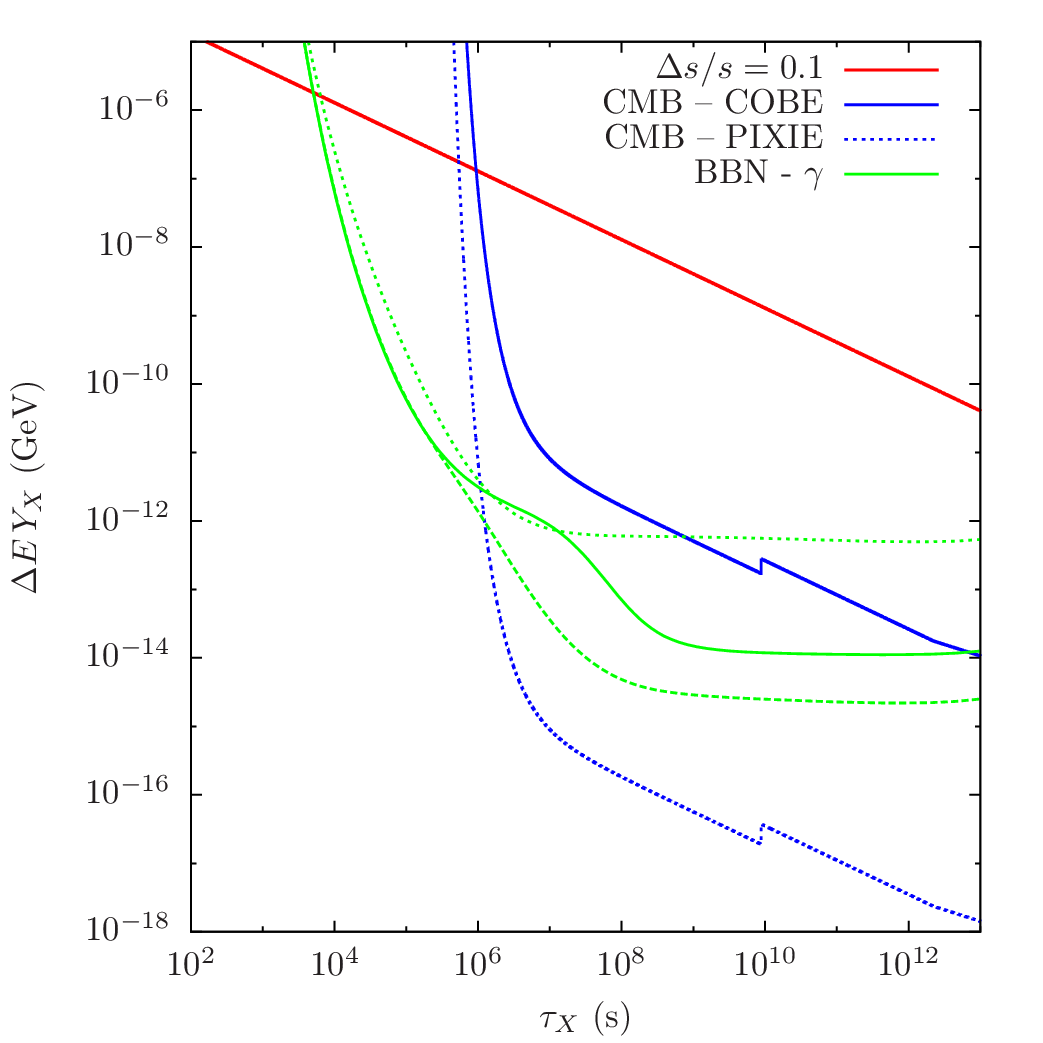}
    \includegraphics[width = 0.45\textwidth]{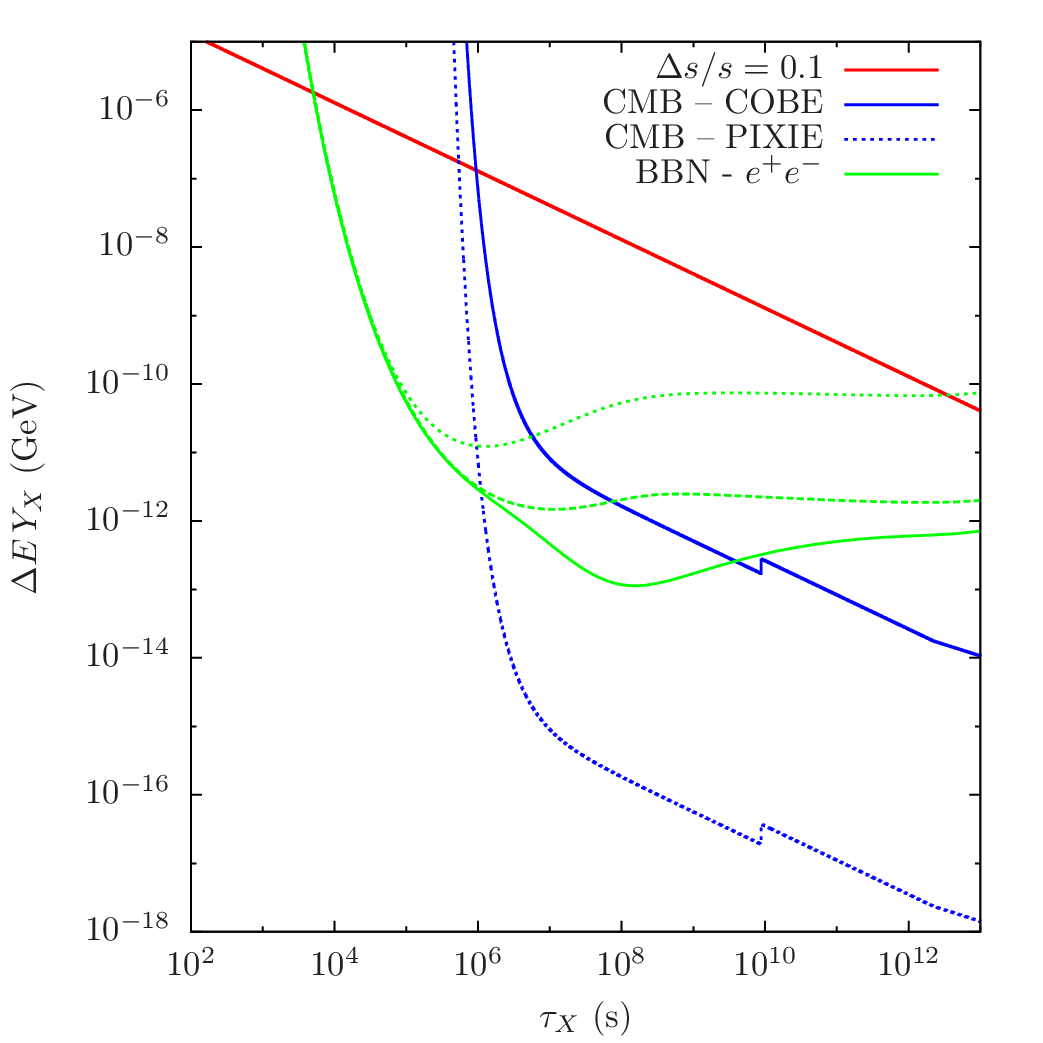}
  \end{center}
\vspace{-0.5cm}
  \caption{
Other bounds on electromagnetic decays in the early universe
as a function of the lifetime $\tau_X$ and the total electromagnetic
injection $\Delta E\,Y_X$ relative to limits derived from BBN.
In both panels, the red line shows $\Delta s/s = 0.1$, while
the solid~(dotted) blue lines show the current and projected
CMB frequency bounds from COBE/FIRAS~(PIXIE).
The left panel also indicates the limits derived from BBN for
photon injection with energy $E_X=10,\,30,\,100\,\mev$ with green dotted, 
dashed, and solid lines.  The right panel shows the corresponding BBN 
bounds from monochromatic $e^+e^-$ injection.
}
  \label{fig:cmbmu}
\end{figure}

  In the left and right panels of Fig.~\ref{fig:cmbmu} we show the limits 
from entropy injection and CMB spectral distortions.  
The solid red line shows $\Delta s/s = 0.1$, and demanding variations 
below this is a conservative requirement relative to those imposed 
in Refs.~\cite{Ishida:2014wqa,Poulin:2015opa}.  
For CMB spectral distortions we show bounds on the $\mu$ and $y$ parameters 
in blue based on the approximate estimates above based on measurements 
by COBE/FIRAS~(solid) and the projected sensitivity of PIXIE~(dotted).
For comparison, we show in green the limits derived above for monochromatic 
photon injection~(left) and monochromatic $e^+e^-$ injection~(right).
In both panels, the dotted, dashed, and solid lines correspond to injection
with $E_X = 10,\,30,\,100\,\mev$.  Even for low injection energies,
BBN constraints currently dominate for $\tau \gtrsim 10^4\,\mathrm{s}$
until being replaced by bounds from either CMB frequency or power spectrum
variations.  Even with the vast improvement expected from PIXIE,
BBN will continue to provide the strongest limit on electromagnetic
decays in the early universe with lifetimes 
$10^4\,\mathrm{s} \lesssim \tau_X \lesssim 10^6\,\mathrm{s}$
and energy injections above a few MeV.

\section{Conclusions\label{sec:conc}}
  
  In this paper we have investigated the electromagnetic cascades induced
by electromagnetic energy injection in the range $E_X = 1\!-\!100\,\mev$
and we have studied its effects on the light elements abundances created
during BBN.  As in Ref.~\cite{Poulin:2015opa}, we find significant
deviations from the universal photon spectrum for monochromatic initial
photon injection with energy $E_X \lesssim E_c = m_e^2/22T$.  We also study 
how this impacts BBN.  Our study also expands on previous work by computing
the full electromagnetic cascade including electrons.  For either photon
or electron injection, we find that BBN provides the strongest constraint
on late-decaying particles with lifetimes between 
$10^4\,\text{s} \lesssim \tau \lesssim 10^{13}\,\text{s}$ for electromagnetic
energies nearly all the way down to the photodissociation threshold of
deuterium near $E_{th} \simeq 2.22,\mev$.

  Photon and electron injection produce very similar electromagnetic cascades
for $E_X \gg E_c$ but differ in important ways for $E_X \lesssim E_c$.
Initial hard photons induce a smooth population of lower-energy photons through
Compton and photon-photon scattering.  In contrast, electrons injected
with $E_X \lesssim m_e^2/10T$ interact mainly through inverse Compton~(IC) 
scattering off the CMB, which lies in the Thomson regime at such energies.  
The upscattered photons from Thomson scattering have much lower energy
than the initial electron, and can easily fall below the MeV scales needed
to induce photodissociation.  However, in this regime we find that
photons radiated off the initial hard electrons can populate and dominate
the induced photon spectrum up to near the initial electron energy.
To our knowledge, the contribution of FSR to the photon spectrum has not been 
considered before in this context since its effects are very small at the
higher initial injection energies that have been investigated 
in the greatest detail.

  While this work has concentrated on decays, our results for electromagnetic
cascades are also applicable to annihilation in the early universe.
Our results could also be used to investigate potential solutions
to the apparent anomalies in the lithium abundances, which was studied
in Refs.~\cite{Ishida:2014wqa,Kusakabe:2013sna} using the universal spectrum.

\bigskip

\begin{flushleft}
\textit{Note added: as this work was nearing completion, Ref.~\cite{Hufnagel:2018bjp} appeared investigating many of the same topics including the development of the electromagnetic cascades from sub-100~MeV decays.  In the cases that are directly comparable, our results appear to be in substantial agreement.
}
\end{flushleft}

\section*{Acknowledgements}

We thank
Sonia Bacca, Nikita Blinov, David Curtin, Barry Davids, David McKeen,
Maxim Pospelov, and Adam Ritz for helpful discussions.
DEM and GW thank the Aspen Center for Physics, which is supported by National 
Science Foundation grant PHY-1607611, for their hospitality
while this work was being completed.
This work is supported by the Natural Sciences
and Engineering Research Council of Canada~(NSERC), with DEM 
supported in part by Discovery Grants and LF by a CGS~D scholarship.
TRIUMF receives federal funding via a contribution agreement 
with the National Research Council of Canada.


\newpage

\appendix

\section{Relaxation and Transfer Rates\label{sec:appa}}

  In this appendix we collect explicit expressions for the relevant
relaxation rates and transfer functions.  The processes included for
each are listed in Table~\ref{tab:proc} together with links to the
corresponding equations.

\begin{table}[ttt]
\beq
\begin{array}{c|c|c}
&~\mathrm{Processes}&~\mathrm{Equations}~\\
\hline
\Gamma_\gamma&\mathrm{4P, PP, PCN, CS}&\eqref{eq:gg4p}, \eqref{eq:ggpp}, \eqref{eq:ggpcn}, \eqref{eq:ggcs}\\
\Gamma_e&\mathrm{IC}&\eqref{eq:geic}\\
K_{\gamma\gamma}&\mathrm{PP, CS}&\eqref{eq:kggpp}, \eqref{eq:kggcs}\\
K_{\gamma e}&\mathrm{IC}&\eqref{eq:kgeic}\\
K_{e\gamma}&\mathrm{4P, PCN, CS}&\eqref{eq:keg4p}, \eqref{eq:kegpcn}, \eqref{eq:kegcs}\\
K_{ee}&\mathrm{IC}&\eqref{eq:keeic}\\
\end{array}
\nnmb
\eeq
\caption{Summary of the contributions to relaxation rates 
and energy transfer functions.
\label{tab:proc}}
\end{table}

\subsection{Rates for Photon Photon Pair Production~(4P)}

  The 4P process plays a key role in determining the electromagnetic
cascade for $E > E_c$ and contributes to $\Gamma_{\gamma}$ 
and $K_{e\gamma}$.  

\subsubsection{4P Photon Relaxation}

The relevant expression is given in Ref.~\cite{Kawasaki:1994sc}
\beq
\Delta\Gamma_{\gamma}(\eg) = 
\frac{1}{8}\frac{1}{\eg^2}\int_{m_e^2/\eg}^{\infty}\!d\bar{E}\,
\frac{1}{\pi^2}\left(e^{\bar{E}/T}-1\right)^{-1}
\int_{4m_e^2}^{4\eg\bar{E}}\!ds\,s\!\!\!\!\!
\left.\phantom{\frac{I}{i}}\sigma_{4P}(\beta)\right|_{\beta = \sqrt{1-4m_e^2/s}} \ ,
\label{eq:gg4p}
\eeq
where
\beq
\sigma_{4P}(\beta) = \frac{\pi}{2}\frac{\alpha^2}{m_e^2}\,
(1-\beta^2)\left[(3-\beta^4)\ln\lrf{1+\beta}{1-\beta}-2\beta(2-\beta^2)\right] \ .
\eeq
This expression can be written in the form
\beq
\Delta\Gamma_{\gamma}(\eg) = \frac{2}{\pi}\alpha^2\lrf{T}{m_e}^3m_e
\frac{1}{y_{\gamma}^2}\,
\mathcal{G}(y_\gamma) \ ,
\eeq
where $y_\gamma = E_\gamma T/m_e^2$ and a function $\mathcal{G}(y_\gamma)$.  
A good numerical fit to this function is
\beq
\frac{1}{y_{\gamma}^2}\mathcal{G}(y_{\gamma}) 
~\simeq~ 
e^{-1/y_\gamma}
\left[(1.2)y_{\gamma}^{-0.425}e^{-(0.1y_{\gamma})^2}
+ (3.3)y_{\gamma}^{-0.85}e^{-(9/y_{\gamma})^2}\right] \ ,
\eeq
which is valid to better than about $10\%$ for $y_{\gamma}\in [10^{-5},10^2]$.
A similar approximate expression is given in Ref.~\cite{Protheroe:1994dt}.

\subsubsection{4P Transfer to Electrons}

  The 4P process also produces high energy electrons and positrons,
contributing to the transfer function $K_{e\gamma}$ by~\cite{Agaronyan:1983xx,Kawasaki:1994sc}:\footnote{This than Ref.~\cite{Agaronyan:1983xx} by a factor of two to account each collision producing an electron and a positron.}
\beq
\Delta K_{e\gamma}(\ee,\eg) = 
\frac{\alpha^2m_e^2}{2\pi}
\frac{1}{\eg^3}\int_{0}^{\infty}\!d\bar{E}_{\gamma}\,
\left(e^{\bar{E}_{\gamma}/T}-1\right)^{-1}\,G(\ee,\eg,\bar{E}_{\gamma})\,\Theta(\{E_i;T\}) \ ,
\label{eq:keg4p}
\eeq
where
\beq
G(\ee,\eg,\bar{E}_{\gamma}) = 4A\,\ln\lrf{4B}{A} - (1-1/B)A^2+2A(2B-1) - 8B \ ,
\eeq
with
\beq
A = \frac{(\eg+\bar{E}_{\gamma})^2}{\ee(\eg+\bar{E}_{\gamma}-\ee)} \ , ~~~~~
B = \frac{\bar{E}_{\gamma}(\eg+\bar{E}_{\gamma})}{m_e^2} \ ,
\eeq
and $\Theta(\{E_i;T\})$ is zero unless $B\geq 1$ and
\beq
\frac{\eg+\bar{E}_{\gamma}}{2}\left(1-\sqrt{1-1/B}\right) 
~\leq~ \ee ~\leq~ \frac{\eg+\bar{E}_{\gamma}}{2}\left(1+\sqrt{1-1/B}\right)  \ .
\eeq
These inequalities reflect kinematic constraints on the 
electron and positron energies.\footnote{These limits were not specified explicitly in Ref.~\cite{Kawasaki:1994sc}.}

\subsection{Rates for Inverse Compton Scattering~(IC)}

  Inverse Compton~(IC) scattering in this context corresponds
to the scattering of high energy electrons (or positrons) off CMB photons,
$e^{\pm}+\gamma_{BG} \to e^{\pm}+\gamma$.  This reduces the electron
energy and produces an energetic photon.  We evaluate the corresponding
contributions to $\Gamma_{e}$, $K_{ee}$, and $K_{\gamma e}$.

\subsubsection{IC Electron Relaxation}

  The relaxation rate for electrons due to IC was computed 
in Ref.~\cite{Jones:1968zza} and reproduced in Ref.~\cite{Kawasaki:1994sc}.
For $e(E_e)+\gamma(\bar{E}_{\gamma}) \to e(E'_e) + \gamma(E_{\gamma})$,
the expression is
\beq
\Delta\Gamma_e(E_e) &=& \frac{2}{\pi}\alpha^2\frac{1}{\ee^2}
\int_0^{\ee}\!d\eg\,
\int_{0}^{\infty}\!d\bar{E}_{\gamma}\;\bar{E}_{\gamma}\big(e^{\bar{E}_{\gamma}/T}-1\big)^{-1}
F(\eg,\ee,\bar{E}_{\gamma}) \ ,
\label{eq:geic}
\eeq
where\footnote{The integration over $\eg$ in Eq.~\eqref{eq:geic}
covers the interval $[0,\ee]$, and not $[\ee,\infty)$ as given
in Ref.~\cite{Kawasaki:1994sc}.}
\beq
F(\eg,\ee,\bar{E}_{\gamma}) = \left\{
\begin{array}{lcl}
2q\ln q+(1+2q)(1-q)
+\frac{(\xi q)^2(1-q)}{2(1+\xi q)}
&;&q\in [0,1]\\
&&\\
0&;&\text{otherwise}
\end{array}
\right.
\label{eq:funcic}
\eeq
with
\beq
\xi = \frac{4\bar{E}_\gamma\ee}{m_e^2} \ ,~~~~~
q = \frac{\eg}{\xi(\ee-\eg)} \ .
\eeq
Note that this expression is based on an approximation that is only
valid for $\bar{E}_{\gamma} < E_{\gamma}$ and 
$q \geq (m_e/2E_e)^2$~\cite{Jones:1968zza}.  However, 
for the energies and temperatures relevant for photodissociation,
we find that applying $q\geq 0$ is a good approximation.
The IC electron relaxation rate can be written in the simpler form
\beq
\Delta\Gamma_e(E_e) = \frac{2}{\pi}\alpha^2\lrf{T}{m_e}^3m_e\,
\frac{1}{y_e^2}\mathcal{H}(y_e) \ ,
\eeq
with
\beq
\frac{1}{y_e^2}\mathcal{H}(y_e) 
&=& \int_0^{y_e}\!dy_{\gamma}\int_0^{\infty}\!dy\,y(e^y-1)^{-1}
F(\eg,\ee,\bar{E}_{\gamma}) \\
&\simeq&
3.07\big/(1+12\,y_e+y_e^2)^{0.47} \ ,
\eeq
where the second line is a numerical fit valid to within about 5\% over
$y_e\in [10^{-5},10^2]$.  For $y_e \ll 1$ this coincides
with IC in the Thomson limit:
$\Delta\Gamma_e = \sigma_T\,n_{\gamma}(T)$, 
where $\sigma_T = (8\pi/3)\alpha^2/m_e^2$ is the Thomson cross section
and $n_{\gamma}$ is the photon density.

\subsubsection{IC Transfer to Electrons}

  The IC process also produces an electron with a lower energy than the initial
value.  The corresponding transfer kernel for 
$e(E_e')+\gamma(\bar{E}_{\gamma}) \to e(E_e) + \gamma(E_{\gamma})$ is
\beq
\Delta K_{ee}(\ee,\ee') = \frac{2}{\pi}\alpha^2\frac{1}{{\ee'}^2}
\int_0^{\infty}\!d\bar{E}_{\gamma}\,\bar{E}_{\gamma}(e^{-\bar{E}_{\gamma}/T}-1)^{-1}
F(\eg,\ee',\bar{E}_{\gamma}) \ , 
\label{eq:keeic}
\eeq
where $\ee' > \ee$,
\beq
\eg = \ee' + \bar{E}_{\gamma} - \ee \ ,
\eeq
and the function $F(\eg,\ee',\bar{E}_{\gamma})$ is given by Eq.~\eqref{eq:funcic}.

\subsubsection{IC Transfer to Photons}

  For photon energy transfer via 
$e(\ee)+\gamma(\bar{E}_{\gamma}) \to e(E_e') + \gamma(\eg)$,
the kernel is\footnote{The corresponding expression 
in Ref.~\cite{Kawasaki:1994sc} appears to have a typo by a factor of two.}
\beq
\Delta K_{\gamma e}(\eg,\ee) &=& \frac{2}{\pi}\alpha^2\frac{1}{\ee^2}
\int_0^{\infty}\!d\bar{E}_{\gamma}\;\bar{E}_{\gamma}(e^{-\bar{E}_{\gamma}/T}-1)^{-1}
F(\eg,\ee,\bar{E}_{\gamma}) 
\label{eq:kgeic}
\eeq
where $F(\{E_i\})$ is defined in Eq.~\eqref{eq:funcic}.

\subsection{Photon Photon Scattering~(PP)}

  Photon photon~(PP) scattering contributes to photon relaxation and transfer.

\subsubsection{PP Photon Relaxation}

The contribution to the relaxation rate
is approximately~\cite{Svensson:1990pfo,Poulin:2015opa}
\beq
\Delta\Gamma_{\gamma}(\eg) ~\simeq~ (0.1513)\alpha^4m_e\lrf{T}{m_e}^3
y_{\gamma}^3\,e^{-y_{\gamma}} \ , 
\label{eq:ggpp}
\eeq  
with $y_{\gamma} = \eg T/m_e^2$.  Note that this form only applies
for $y_{\gamma} < 1$, and thus we add an exponential by hand to 
provide a smooth cutoff at larger values.

\subsubsection{PP Photon Transfer}

For transfer $\gamma(\eg')+\gamma_{BG} \to \gamma(\eg)+\gamma$,
the kernel is~\cite{Svensson:1990pfo,Poulin:2015opa}
\beq
\Delta K_{\gamma\gamma}(\eg,\eg') ~\simeq~
(0.4324)\alpha^4\lrf{T}{m_e}^4{y_{\gamma}'}^2\left(
1-\frac{y_{\gamma}}{y_{\gamma}'}+\frac{y_{\gamma}^2}{{y_{\gamma}'}^2}
\right)^2e^{-y_{\gamma}} \ ,
\label{eq:kggpp}
\eeq
with $\eg < \eg'$ and an exponential has again been added to cut off this 
form when $y_\gamma > 1$.

\subsection{Pair Creation on Nuclei~(PCN)}

  Photon scattering on background nuclei can create $e^+e^-$ pairs,
$\gamma(\eg)+ N \to N + e^++e^-$.  This reduces the photon energy
and injects energy into electrons and positrons.

\subsubsection{PCN Photon Relaxation}

  The contribution to photon relaxation is~\cite{Kawasaki:1994sc}
\beq
\Delta\Gamma_{\gamma}(\eg) = \sum_Zn_Z\sigma_{PCN}^{(Z)}(\eg)
\label{eq:ggpcn}
\eeq
where the sum runs over $Z = \mathrm{H},\,{}^4\mathrm{He}$
and $n_Z$ are their corresponding densities.  
The cross section for $x = 2\eg/m_e \geq 8$ is
\beq
\sigma_{PCN}^{(Z)} &=& Z^2\frac{\alpha^3}{m_e^2}
\left(\frac{28}{9}\ln x - \frac{218}{27}  \right.\\
&&
~~+ \left.\lrf{4}{x}^2\left[\frac{2}{3}(\ln x)^3 - (\ln x)^2
+\big(6-\frac{\pi^2}{3}\big)\ln x + 2\zeta(3) + \frac{\pi^2}{6} - \frac{7}{2}
\right]
\right)+ \ldots
\nnmb
\eeq 
while for $4 \leq x < 8$ it is
\beq
\sigma_{PCN}^{(Z)} = Z^2\frac{2\pi}{3}\frac{\alpha^3}{m_e^2}
\lrf{x-4}{x}^3\left(1+ \frac{1}{2}\rho+\frac{23}{40}\rho^2  
+\frac{11}{60}\rho^3 + \frac{29}{960}\rho^4
+ \ldots\right) \ ,
\eeq
with 
\beq
\rho = \frac{x-4}{2+2\sqrt{x}+x/2} \ .
\eeq
Note that $x=4$ is the threshold for the process to occur.

\subsubsection{PCN Electron Transfer}

  The contribution to the electron transfer kernel is~\cite{Kawasaki:1994sc}
\beq
\Delta K_{e\gamma }(\ee,\eg) = \sum_Zn_Z\frac{d\sigma_{PCN}^{(Z)}}{d\ee} \ ,
\label{eq:kegpcn}
\eeq
with 
\beq
\frac{d\sigma_{PCN}^{(Z)}}{d\ee} &=&
Z^2\frac{\alpha^3}{m_e^2}\lrf{pp'}{\eg^3}\Bigg(\\
&&~
-\frac{4}{3}-2\ee\ee'\frac{p^2+{p'}^2}{p^2{p'}^2}
\nnmb\\
&&
~~+m_e^2\left(\frac{\ell'\ee}{{p'}^3}+\frac{\ell\ee'}{p^3}
-\frac{\ell\ell'}{pp'}\right)
\nnmb\\
&& ~~~+L\bigg[
-\frac{8\ee\ee'}{3pp'} + \frac{\eg^2}{p^3{p'}^3}\left(
\ee^2{\ee'}^2+p^2{p'}^2-m_e^2\ee\ee'\right)
\nnmb\\
&&~~~~-\frac{m_e^2\eg}{2pp'}\left(\ell\frac{\ee\ee'-p^2}{p^3}
+\ell'\frac{\ee\ee'-{p'}^2}{{p'}^3}\right)
\bigg]
\Bigg)
\nnmb
\eeq
where $\ee' = (\eg - \ee)$, $p^{(')} = \sqrt{{\ee^{(')\,2}}-m_e^2}$, and 
\beq
\ell^{(')} &=& \ln\bigg(\frac{\ee^{(')}+p^{(')}}{\ee^{(')}-p^{(')}}\bigg) \ ,\\
L &=& \ln\lrf{\ee\ee'+pp'+m_e^2}{\ee\ee'-pp'+m_e^2} \ .
\eeq
As before, the relevant nuclei are H and $^4$He.

\subsection{Compton Scattering~(CS)}

  Compton scattering in this context refers to high-energy photons
colliding with background electrons, $\gamma + e^-_{BG} \to \gamma + e^-$.
This reduces the photon energy and transfers it to the scattered electron.

\subsubsection{CS Photon Relaxation}

The contribution to the relaxation rate is
\beq
\Delta\Gamma_{\gamma}(\eg) = n_e\sigma_{CS} \ ,
\label{eq:ggcs}
\eeq
where $n_e$ is the background electron density given and
\beq
\sigma_{CS}(\eg) = 2\pi\frac{\alpha^2}{m_e^2}\,\frac{1}{x}
\left[\left(1-\frac{4}{x}-\frac{8}{x^2}\right)\ln(1+x)+\frac{1}{2}
+\frac{8}{x}-\frac{1}{2(1+x)^2}\right] \ ,
\eeq
with $x = 2\eg/m_e$.

\subsubsection{CS Photon Transfer}

  The energy of the initial CS photon $\eg'$ is partially transferred to
the energy $\eg$ of the outgoing photon.  This contributes to the
transfer kernel by~\cite{Kawasaki:1994sc,Poulin:2015opa,Slatyer:2009yq}
\beq
\Delta K_{\gamma\gamma}(\eg,\eg') = 
n_e\,\frac{d\sigma_{CS}(\eg,\eg')}{d\eg}
\label{eq:kggcs}
\eeq
with
\beq
\frac{d\sigma_{CS}(\eg,\eg')}{d\eg} = 
\pi\frac{\alpha^2}{m_e}\,
\frac{1}{{\eg'}^2}\left[\frac{\eg'}{\eg}+\frac{\eg}{\eg'}
+\left(\frac{m_e}{\eg}-\frac{m_e}{\eg'}\right)^2
- 2m_e\left(\frac{1}{\eg}-\frac{1}{\eg'}\right)\right] \ .
\label{eq:dsigcs}
\eeq
This expression is valid only for $\eg'/(1+2\eg'/m_e) \leq \eg \leq \eg'$
and zero otherwise.\footnote{
The differential CS expression in Ref.~\cite{Poulin:2015opa}
appears to have an incorrect sign in the last term of Eq.~\eqref{eq:dsigcs},
and neither Refs.~\cite{Kawasaki:1994sc,Poulin:2015opa} give the lower bound
on $\eg$ stated explicitly in Ref.~\cite{Slatyer:2009yq}.}

\subsubsection{CS Electron Transfer}

  Energy transfer to electrons by CS is given by
\beq
\Delta K_{e\gamma}(\ee,\eg') = 
n_e\,\frac{d\sigma_{CS}(\eg,\eg')}{d\eg} \ ,
\label{eq:kegcs}
\eeq
where $\eg = (\eg'+m_e-\ee$), $\eg' \geq \ee$, and
the differential cross section is given in Eq.~\eqref{eq:dsigcs}.

\bibliographystyle{JHEP}
\bibliography{ref_bbnem}

\end{document}